\documentclass[prd,aps,preprint,tightenlines,nofootinbib,superscriptaddress]{revtex4}
\usepackage{mathrsfs}
\usepackage{amsfonts}
\usepackage{amsmath}
\usepackage{array}
\usepackage{verbatim}
\usepackage{epsfig}
\usepackage{graphicx}
\usepackage{epstopdf}
\usepackage{graphicx}
\usepackage{ulem}
\usepackage{color}

\begin{document}
\title{Universal Non-perturbative Functions for SIDIS and Drell-Yan Processes}

\author{Peng Sun}
\affiliation{Nuclear Science Division, Lawrence Berkeley National
Laboratory, Berkeley, CA 94720, USA}
\author{Joshua Isaacson}
\affiliation{Department of Physics and Astronomy, Michigan State University,
East Lansing, MI 48824, USA}
\author{C.-P. Yuan}
\affiliation{Department of Physics and Astronomy, Michigan State University,
East Lansing, MI 48824, USA}
\author{Feng Yuan}
\affiliation{Nuclear Science Division, Lawrence Berkeley National
Laboratory, Berkeley, CA 94720, USA}

\begin{abstract}
We update the well-known BLNY fit to the low transverse momentum Drell-Yan lepton pair
productions in hadronic collisions, by considering the constraints from the semi-inclusive
hadron production in deep inelastic scattering (SIDIS) from HERMES and
COMPASS experiments. We follow the Collins-Soper-Sterman (CSS) formalism
with the $b_*$-prescription. A universal non-perturbative form factor
associated with the transverse momentum dependent quark distributions
is found in the analysis with a new functional form different from that of
BLNY. This releases the tension between the BLNY fit to the Drell-Yan data
with the SIDIS data from HERMES/COMPASS in the CSS resummation formalism.
\end{abstract}

 \maketitle

\section{introduction}

To reliably predict the transverse momentum distribution of the final state particles
in some scattering processes in hadron collisions may require all
order resummation of large logarithms.
Among these hard processes, two of the classic
examples include the Drell-Yan lepton pair production
and the semi-inclusive hadron production in
deep inelastic scattering (SIDIS)~\cite{Boer:2011fh}. In both processes, there are two
separate scales: the virtuality of the virtual photon $Q$ and
the transverse momentum of either final state virtual photon $q_\perp$
in Drell-Yan process or final state hadron $P_{h\perp}$ in DIS process.
Large logarithms exist in higher order perturbative calculations when
$Q$ is much larger than $q_\perp$ and are of the form:
$\alpha_s^i\left(\ln Q^2/q_\perp^2\right)^{2i-1}$~\cite{Collins:1981uk,Collins:1984kg,Dokshitzer:1978dr,Parisi:1979se}.
The resummation of these
large logarithms are carried out by applying the transverse
momentum dependent (TMD) factorization
and evolutions~\cite{Collins:1981uk,Collins:1984kg,Ji:2004wu,Collins:2004nx,Collins},
where the non-perturbative form factors associated
with the TMD parton distributions play an
important role~\cite{Landry:2002ix,Konychev:2005iy,Qiu:2000ga,Kulesza:2002rh,Catani:2000vq,
Catani:2003zt,Bozzi:2003jy}. This resummation is usually referred to as
the TMD resummation or Collins-Soper-Sterman (CSS)
resummation. Following the QCD
factorization arguments and the universality of the TMD
parton distributions, we shall expect that the non-perturbative
functions determined from Drell-Yan processes can be applied to the SIDIS
processes as well,
subject to the needed modification for taking into account the
fragmentation function contribution in order to generate
final state transverse momentum distributions.
Recent experimental measurements of SIDIS processes
from the HERMES~\cite{Airapetian:2012ki} and COMPASS~\cite{Adolph:2013stb} collaborations provide an
opportunity to understand the TMD distributions in both processes,
which have already attracted several theory studies~\cite{Sun:2013dya,Signori:2013mda,Anselmino:2013lza,Aidala:2014hva,Echevarria:2014xaa}.
The goal of the current paper is to investigate the universality of the TMD parton distributions
in the CSS resummation formalism to simultaneously describe the transverse momentum
distributions in the Drell-Yan and SIDIS processes~\footnote{The SIDIS processes
in the very small-$x$ region from HERA measurements have been analyzed
in Ref.~\cite{Nadolsky:1999kb} in the CSS resummation, where a totally different functional
form has been used to describe the experimental data. Since the HERA data
covers mostly the small-$x$ region, we will come back to them in a future publication.}.

We will start with the well-known Brock-Landry-Nadolsky-Yuan (BLNY)  fit to the transverse momentum
dependent Drell-Yan lepton pair productions in hadronic collisions~\cite{Landry:2002ix}.
The BLNY fit parameterizes the non-perturbative form factors as
$\left(g_1+g_2\ln (Q/2Q_0)+g_1g_3\ln(100\,x_1x_2)\right))b^2$
in the impact parameter space with $x_1$ and $x_2$ representing the longitudinal
momentum fractions of the incoming nucleons carried by
the initial state quark and antiquark. These parameters are
constrained from the combined fit to the low transverse momentum
distributions of  Drell-Yan lepton pair production
with $4\textmd{GeV} < Q < 12\textmd{GeV}$ in fixed
target experiments and $Z$ production ($Q\sim 90\textmd{GeV}$)
at the Tevatron. These results can also be applied to $W$ production at the Tevatron.
However, this parameterization does not apply to the SIDIS processes
measured by HERMES and COMPASS collaborations: if we extrapolate the above parameterization
down to the typical HERMES kinematics
where $Q^2$ is around $ 3\textmd{GeV}^2$, we can not describe the
transverse momentum distribution of hadron production
in the experiments~\cite{Sun:2013dya}.

In this paper, we provide a novel parametrization form to consistently describe the Drell-Yan
data and SIDIS data in the CSS resummation formalism with a universal non-perturbative
TMD function.
In order to describe the SIDIS data, it is necessary to modify the original
BLNY parameterization. In the original BLNY parameterization, there is a strong
correlation between the $x$ and the $Q^2$ dependence
of the non-perturbative form factor~\cite{Landry:2002ix}. This
is because $x_1x_2=Q^2/S$ where $S$ is the
square of the center-of-mass energy of the incoming hadrons.
Therefore, at the first step, we will separate out the $x$-dependence,
and assume a power law behavior: $(x_0/x)^\lambda$.
These two parameterizations (logarithmic and power law) differ
strongly in the intermediate $x$ range.
Second, we modify the $\ln Q$ term in the non-perturbative
form factor by following the observation of Ref.~\cite{Sun:2013dya},
which has shown that a direct integration of the evolution kernel
can describe the SIDIS and Drell-Yan data  with $Q$ values ranging from a few
to ten GeV. Direct integration of the evolution kernel leads to
a functional form of $\ln(b/b_{*})\ln(Q)$, instead of $b^2\ln(Q^2)$.
Therefore, we will
perform a global fit to the selected set of experimental data
with the non-perturbative function:
\begin{equation}
g_1b^2+g_2\ln(b/b_*)\ln(Q/Q_0)+g_3b^2\left((x_0/x_1)^\lambda+(x_0/x_2)^\lambda\right)\ ,
\end{equation}
with $b_*$ defined as,
\begin{equation}
 b_*=b/\sqrt{1+b^2/b_{max}^2}  \ ,~~b_{max}<1/\Lambda_{QCD}\ .\label{bstar}
\end{equation}
After obtaining the TMD non-perturbative function from the fit to the
Drell-Yan data, we apply the fit to the transverse momentum distributions
in SIDIS processes from HERMES and COMPASS,
which also depend on the final state fragmentation functions. We find that the
new parametrization form can describe well the SIDIS data, and
therefore establish the universality property of the TMD
distributions between  DIS
and Drell-Yan processes.

The rest of the paper is organized as follows. In Sec.II, we present
the theoretical framework of the CSS formalism and the basic set-up
in the calculations of the transverse momentum distributions in Drell-Yan
lepton pair production and SIDIS processes. In Sec.III, we
perform a global fit to the
Drell-Yan data with the modified BLNY parameterization.
In Sec. IV, we apply the newly determined non-perturbative function
to the SIDIS processes and demonstrate that it can consistently describe the
transverse momentum distribution measurements from HERMES and
COMPASS Collaborations. We will also comment on the role of
the $Y$-terms in SIDIS at the energy range of HERMES and COMPASS.
Finally, we conclude our paper and comment on the impact
of the new fit.

\section{Collins-Soper-Sterman Formalism for Low Transverse Momentum Drell-Yan
and SIDIS Processes}

In this section, we review the basic formulas of the CSS resummation formalism
and the theory framework to calculate the transverse momentum distributions for the
Drell-Yan lepton pair production at hadron colliders and semi-inclusive
hadron production in DIS processes.
In the (low energy) Drell-Yan lepton pair production in hadronic collisions, we have
\begin{equation}
A (P_A)+B(P_B) \to \gamma^* (q) +X \to \ell^+ + \ell^ -
+X,
\end{equation}
where $P_A$ and $P_B$ represent the momenta of hadrons $A$ and $B$,
respectively.
According to the CSS resummation formalism, the
differential cross section can be expressed as
\begin{eqnarray}
\frac{d^4\sigma}{dQ^2dyd^2q_\perp}&=&
\sigma_0^{\rm (DY)}\int \frac{d^2b}{(2\pi)^{2}} e^{i \vec{q}_\perp\cdot \vec{b}} \widetilde
{W}_{UU}(Q;b) +Y^{\rm (DY)}_{UU}(Q;q_\perp)\ ,
\end{eqnarray}
where $q_\perp$ and $y$ are transverse momentum and rapidity of the lepton pair, respectively,
$\sigma_0^{\rm (DY)}=4\pi\alpha_{em}^2/(3N_cSQ^2)$ with
the color factor $N_c=3$ and $S=(P_A+P_B)^2$.
In the above equation, the first term is dominant in the $q_\perp\ll Q$ region, while the
second term is dominant in the region of $q_\perp\sim Q$ and $q_\perp > Q$.
In this paper, we focus
on the low transverse momentum region to constrain the
non-perturbative form factors, which is embedded in the
first term of the above equation.

Similarly, in the SIDIS process, we have,
\begin{equation}
e (\ell)+p(P)\to e(\ell') + h
(P_h) + X\ ,
\end{equation}
which proceeds through exchange of a virtual photon
with momentum $q_\mu=\ell_\mu-\ell'_\mu$, and invariant mass
$Q^2=-q^2$.  The
differential  SIDIS cross section is written as
\begin{eqnarray}
    \frac{d^5\sigma}{dx_Bdydz_hd^2\vec {P}_{h\perp}}
      &=& \sigma_0^{\rm (DIS)}\frac{1}{z_h^2}\int \frac{d^2b}{(2\pi)^{2}} e^{i \vec{P}_{h\perp}\cdot \vec{b}/z_h} \widetilde
{F}_{UU}(Q;b) +Y^{(DIS)}_{UU}(Q;P_{h\perp}) \ ,
      \end{eqnarray}
where $\sigma_0^{\rm (DIS)}=4\pi\alpha^2_{\rm em}S_{ep}/{Q^4}\times (1-y+y^2/2)
x_B$ with usual DIS kinematic variables $y$, $x_B$, $Q^2$, and $S_{ep}=({\ell}+P)^2$ .
Here, $z_h=P_h\cdot P/q\cdot P$, which denotes
the momentum fraction of the virtual photon carried by the final state hadron.
The transverse momentum of the final state hadron
$P_{h\perp}$ is defined in the lepton-proton center-of-mass frame.

Following the resummation and evolution of these hard processes, we can
write down the following expressions for the
cross sections in the impact parameter space,
\begin{eqnarray}
\widetilde {W}_{UU}(Q;b)&=&e^{-{\cal S}_{pert}(Q^2,b_*)-S_{NP}(Q,b)}
\nonumber\\
&&\times \Sigma_{i,j} C_{qi}^{(DY)}\otimes f_{i/A}(x_1,\mu=b_0/b_*) C_{\bar qj}^{(DY)}\otimes f_{j/B}(x_2,\mu=b_0/b_*) \ ,\\
\widetilde {F}_{UU}(Q;b)&=&e^{-{\cal S}_{pert}(Q^2,b_*)-S_{NP}(Q,b)}\nonumber\\
&&\times \Sigma_{i,j} C_{qi}^{(DIS)}\otimes f_{i/A}(x_B,\mu=b_0/b_*) \hat{C}_{qj}^{(DIS)}\otimes D_{h/j}(z_h,\mu=b_0/b_*) \ ,
\end{eqnarray}
where $b_0=2e^{-\gamma_E}$ with $\gamma_E$ the Euler constant,
$x_{1,2}=Qe^{\pm y}/\sqrt{s}$ represent the momentum fractions carried by the incoming
quark and antiquark in the Drell-Yan processes, $f_{i/A}$ and $D_{h/j}$ denote
the relevant longitudinal parton distribution and
fragmentation functions, respectively. In the above equation,
$b_*$-prescription is introduced~\cite{Collins:1984kg} and $b_*$ follows the definition
in Eq.~(\ref{bstar}).
The perturbative Sudakov form factor resums the large double logarithms
of all order gluon radiation,
\begin{equation}
S_{pert}(Q,b)=\int_{b_0/b}^Q\frac{d\bar\mu}{\bar\mu}\left[A\ln\frac{Q^2}{\bar\mu^2}+B\right] \ ,\label{spert}
\end{equation}
where $A$ and $B$ are calculable order by order in perturbation theory. In the following
numerical calculations, we keep $A$ and $B$ up to 2-loop and 1-loop order, 
respectively, in the QCD interaction.
Meanwhile, we will keep $C$ coefficients and $Y$ terms at one-loop order in the
numerical calculation.

In addition, the $b_*$-prescription in the CSS resummation formalism
introduces a non-perturbative form factor,
and a generic form was suggested~\cite{Collins:1984kg},
\begin{equation}
S_{NP}=g_2(b)\ln Q/Q_0 +g_1(b) \ .
\end{equation}
Here, $g_1$ and $g_2$ are functions of the impact parameter $b$ and they
also depend on the choice of $b_{max}$. In the literature, these functions have been
assumed Gaussian forms for simplicity, i.e., $g_{1,2}\propto b^2$. The most successful
approach is the so-called BLNY parameterization mentioned in the Introduction, which has been encoded in
ResBos program~\cite{Landry:2002ix} with successful applications for vector boson production at
the Tevatron and LHC.
We notice that the above adaption is not the only choice to apply to the CSS
resummation~\cite{Qiu:2000ga,Kulesza:2002rh,Catani:2000vq,
Catani:2003zt,Bozzi:2003jy}.

\section{Update the BLNY Fit for Vector Boson Production in Hadronic Collisions}

In the BLNY fit, the flavor dependence of $S_{NP}$ has been ignored for 
simplicity, and the following functional form has been chosen,
\begin{equation}
S_{NP}=g_1b^2+g_2b^2\ln\left({Q}/{3.2}\right)+g_1g_3b^2\ln(100 x_1x_2)  \ ,\label{blny0}
\end{equation}
for Drell-Yan type of processes in hadronic collisions, where $g_{1,2,3}$ are fitting parameters~\cite{Landry:2002ix},
\begin{equation}
g_1=0.21,~~g_2=0.68,~~g_1g_3=-0.12,~~{\rm with}~~b_{max}=0.5\textmd{GeV}^{-1} \ .
\end{equation}
Although the above parameterizations describe very well the Drell-Yan type of
processes in hadronic collisions from fixed target experiments to
colliders, we can not use them to describe the transverse momentum
distributions of semi-inclusive hadron production in DIS processes, as explained
in great detail in Ref.~\cite{Sun:2013dya}.

The $\ln(Q)$ dependence of $S_{NP}$, cf. Eq.~(\ref{blny0}),  
follows from renormalization-group invariance of soft-gluon radiation, 
and needs to be modified in order to  simultaneously 
describe the Drell-Yan and SIDIS processes using the TMD formalism. 
For that, we follow the observation made in 
Ref.~\cite{Sun:2013dya} that the $g_2$ function should have logarithmic dependence
on $b$, instead of $b^2$ dependence. 
Therefore, in this work, we consider the following parameterization,
\begin{equation}
g_2\ln \left(b/b_*\right) \ln(Q/Q_0) \ .
\end{equation}
At small-$b$, the above function reduces to power behavior as $b^2$, which is
consistent to the power counting analysis in Ref.~\cite{Korchemsky:1994is}. However,
at large $b$, the logarithmic behavior will lead to different predictions
depending on $Q^2$.
It is interesting to note that the above form has been suggested in an earlier
paper by Collins and Soper~\cite{Collins:1985xx}, but has not yet 
been adopted in any phenomenological study.

In addition, we will modify the $x$-dependence in the non-perturbative function
as mentioned in the Introduction so that
\begin{equation}
S_{NP}=g_1b^2+g_2\ln\left(b/b_*\right)\ln\left({Q}/{Q_0}\right)+g_3b^2\left((x_0/x_1)^\lambda+(x_0/x_2)^\lambda\right)
\ ,\label{syy}
\end{equation}
where we have fixed $Q_0^2=2.4\,{\rm GeV}^2$, $x_0=0.01$\footnote{
The choice of $x_0=0.01$ is motivated by the so-called saturation model, 
in which it was assumed that gluon distribution (or quark distribution) 
has saturation behavior as $x<0.01$~\cite{saturation}. 
}
 and $\lambda=0.2$.
The specific $x$-dependence is motivated by some saturation model 
of parton distributions at the small-$x$ values~\cite{saturation}. This functional form
also has mild dependence on $x$ in the intermediate $x$-range as
compared to the original BLNY parameterization 
(with pure Gaussian form in $b$ space).

In the above parameterization, we have chosen $Q_0^2=2.4\,{\rm GeV}^2$ in order to make
it convenient to compare to the final state hadron distribution in SIDIS
experiments from HERMES and COMPASS Collaborations.
From this choice of $Q_0^2$, the
importance of $g_1$ and $g_3$ in SIDIS is clearly illustrated.

\begin{table}[t]
\caption{The non-perturbative functions parameters fitting results. Here,
\protect$ N_{fit}\protect$ is the fitted normalization factor for each
experiment.  } {\centering
\begin{tabular}{|c|c|} \hline Parameter &
 SIYY fit \\
\hline \hline $ g_{1} $ &
0.212\\
\hline $ g_{2} $ &
 0.84\\
\hline $ g_{3}$  &
0.0\\
\hline \hline E288 &
 $ N_{fit}=0.83$  \\ (28 points)
 & $\chi^2=51$
\\
\hline \hline E605 &
 $ N_{fit}=0.85$  \\ (35 points)
 & $\chi^2=60$
\\
\hline \hline R209 &
 $ N_{fit}=1.02$  \\ (10 points)
 & $\chi^2=3$
\\
\hline CDF Run I&
$ N_{fit}= 1.07$ \\ (20 points)
  & $\chi^2=11$
\\
\hline D0 Run I&
$ N_{fit}= 0.94$ \\ (10 points)
  & $\chi^2=8$
\\
\hline CDF Run II&
$ N_{fit}= 1.08$ \\ (29 points)
  & $\chi^2=31$
\\
\hline D0 Run II&
$ N_{fit}= 1.02$ \\ (8 points)
  & $\chi^2=5.3$
\\
\hline \hline $ \chi ^{2} $&
 168.4  \\
 \hline \hline $ \chi ^{2} /$DOF &
 1.26  \\
\hline
\end{tabular}\par}
{\centering \label{fit_result}\par}
\end{table}

Some comments shall follow before we present the result of our analysis.
Firstly, $g_1$ and $g_2$ are generally non-perturbative
functions of $b$ and $x$. We could guess for their 
functional forms, but only experimental data can tell which of
these forms is correct~\footnote{Recent proposal
of a lattice formulation of the TMD parton distributions in Euclidean space may help
to solve this issue in the future~\cite{Ji:2013dva}. Some lattice calculation
attempts can be found in Ref.~\cite{Hagler:2009mb}.}.
Hence, it is important to perform a global fit to the existing 
experimental data to test out the proposed 
non-perturbative function forms. 
In addition to the pure Gaussian form as adopted in the BLNY fit, and 
the logarithmic dependence form as proposed in this work,   
another choice of the non-perturbative form has also been suggested, 
such as the Qiu-Zhang prescription in Ref.~\cite{Qiu:2000ga}.
To discriminate various forms of the non-perturbative function 
$S_{NP}$ would require more precise experimental data than what we 
have at hand. Secondly, we know that $S_{NP}$ shall follow $b^2$ power law 
at small-$b$ values, as given by the 
power counting analysis~\cite{Korchemsky:1994is}.
This requirement imposes a strong constraint to the proposed 
non-perturbative models, and the model we proposed above 
satisfies this constraint.
Most importantly, after fitting to the experimental data, the TMD evolution
shall predict relevant scale dependence for various interesting
observables. For example, the single transverse spin azimuthal asymmetries
will be able to provide additional constraints on the evolution 
of partons in the TMD formalism~\cite{Sun:2013dya}.
This will become possible in the near future with high precision
data from JLab 12 GeV upgrade and the planned electron-ion
collider~\cite{Boer:2011fh}.
In summary, introducing the logarithmic $b$ dependence 
in the $\ln(Q)$ term and the mild $x$-dependence 
in the intermediate-$x$ region, as described in Eq.(\ref{syy}), we are
able to consistently describe the transverse momentum distributions 
in both the Drell-Yan and SIDIS data.

\begin{figure}[tbp]
\centering
\includegraphics[width=7cm]{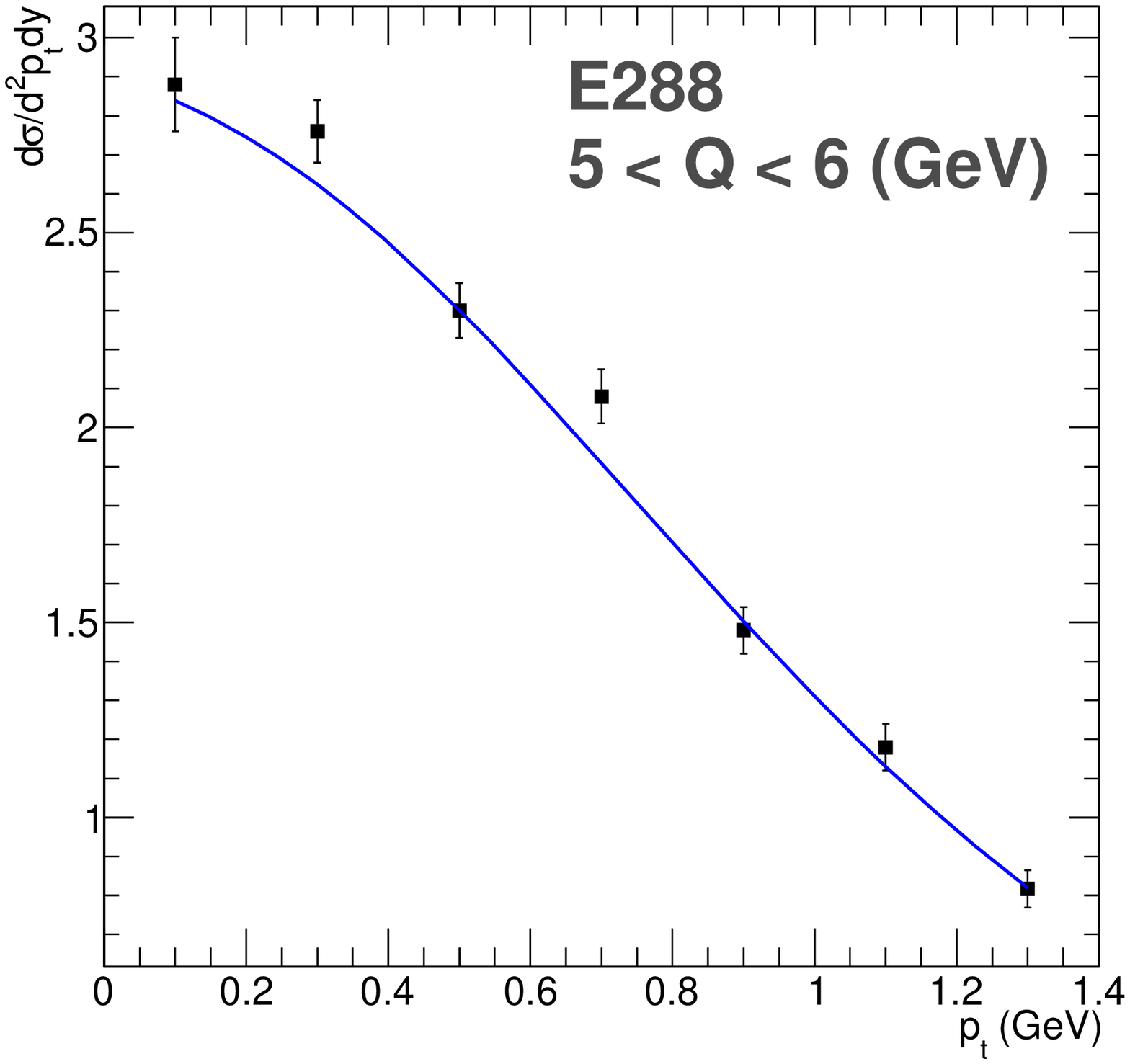}
\includegraphics[width=7cm]{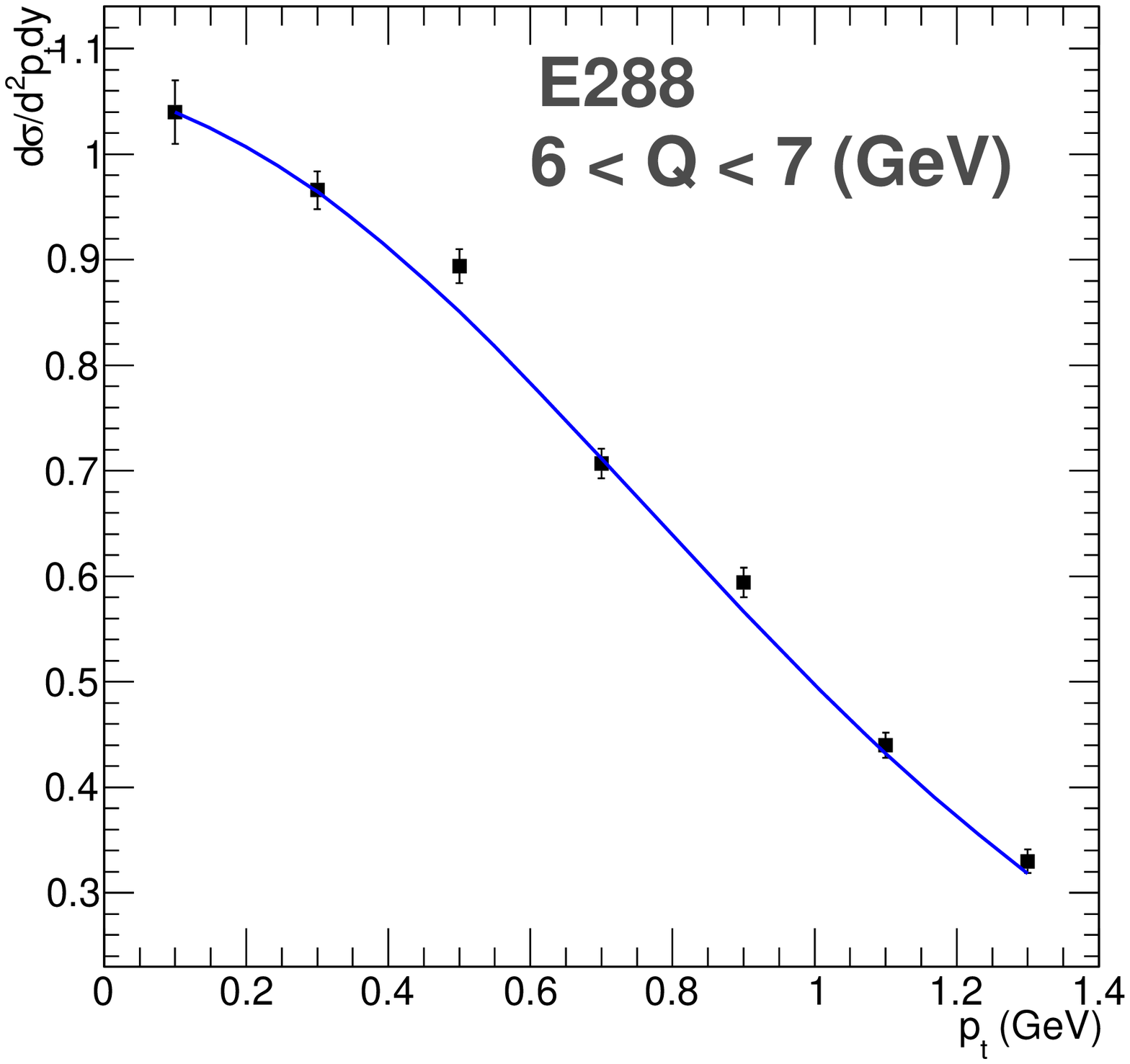}
\includegraphics[width=7cm]{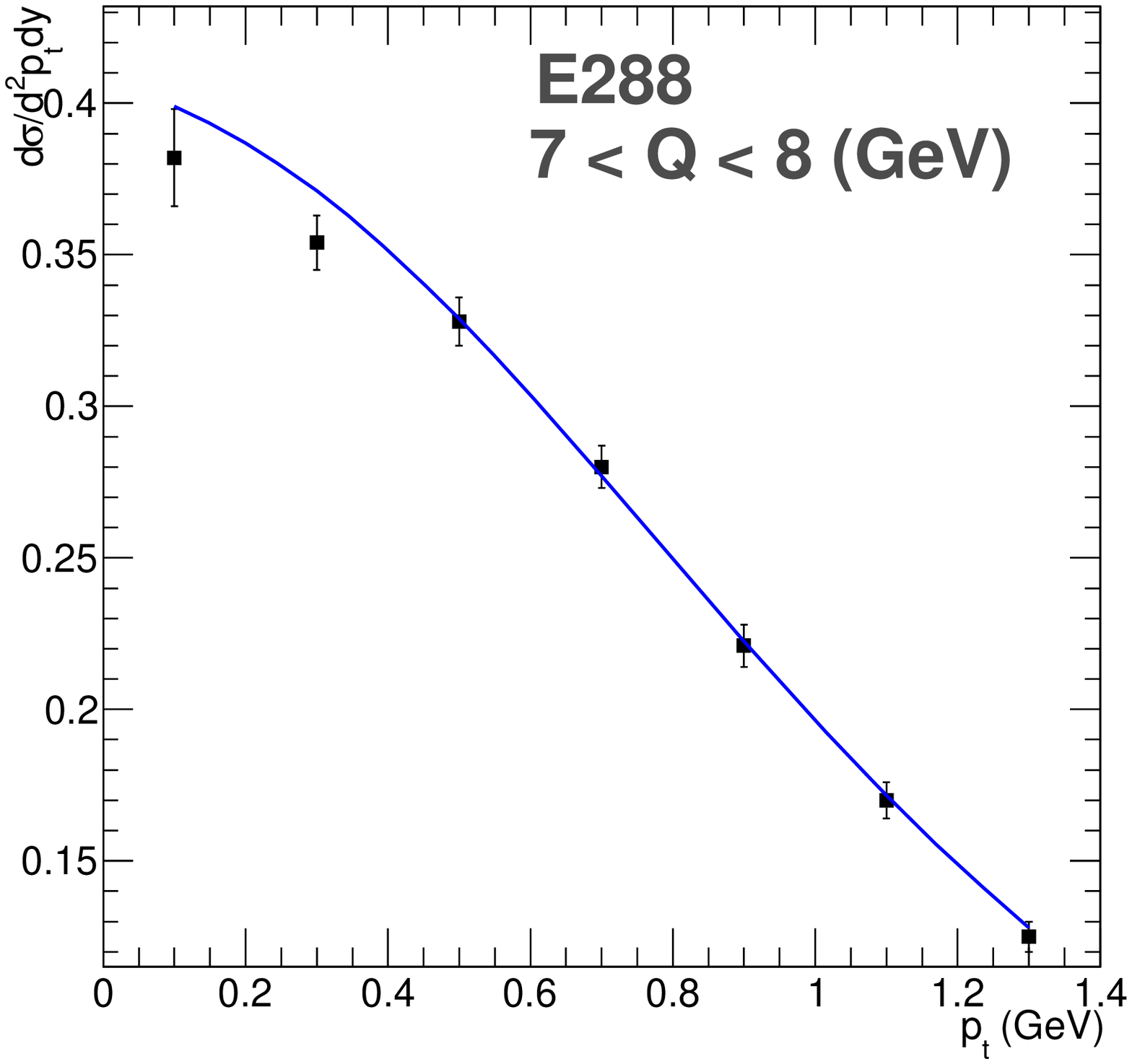}
\includegraphics[width=7cm]{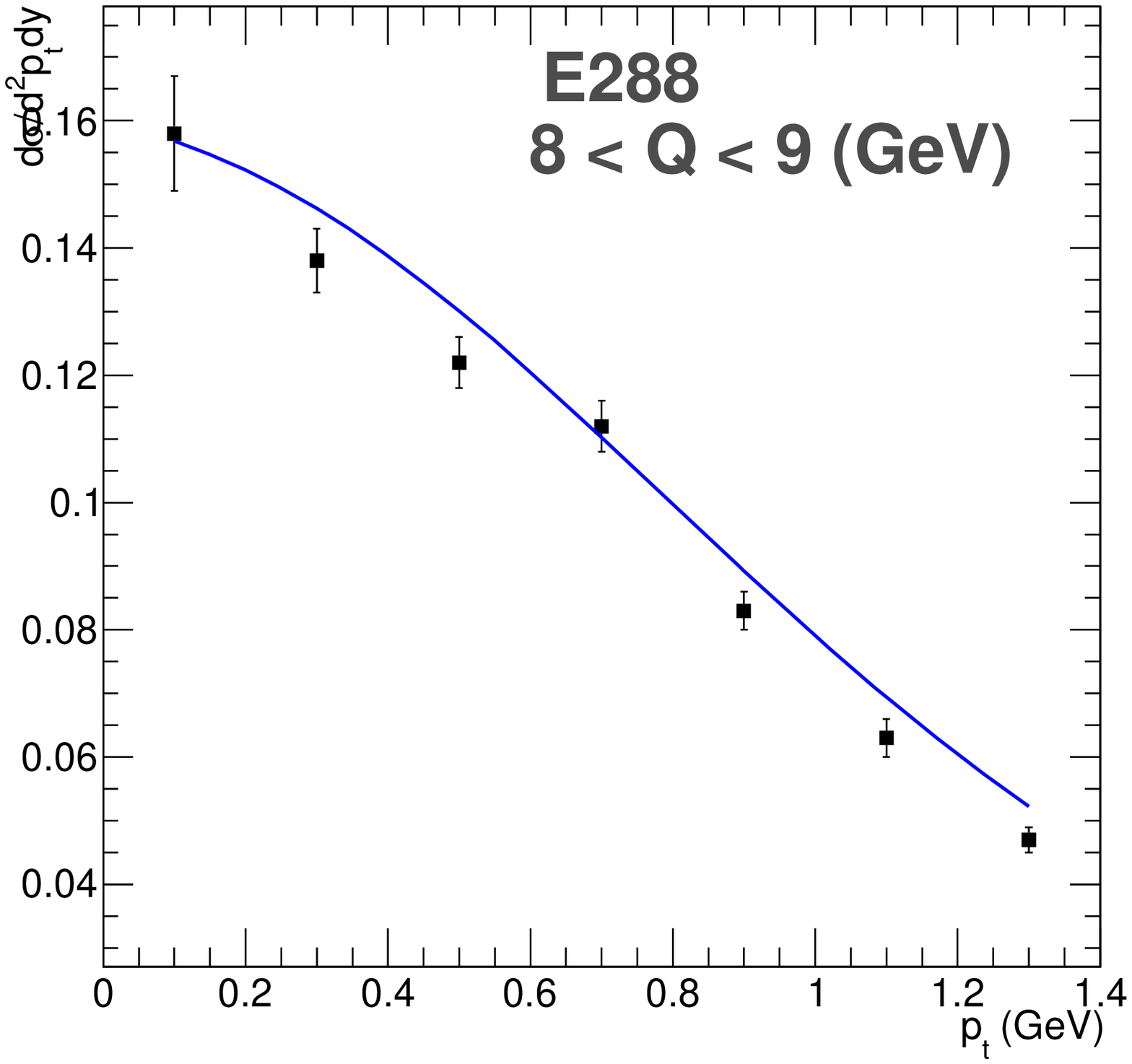}
\caption{Fit to the differential cross section for Drell-Yan lepton pair production
in hadronic collisions from E288 Collaboration~\cite{Ito:1980ev}.}
\label{fig:e288}
\end{figure}

\begin{figure}[tbp]
\centering
\includegraphics[width=7cm]{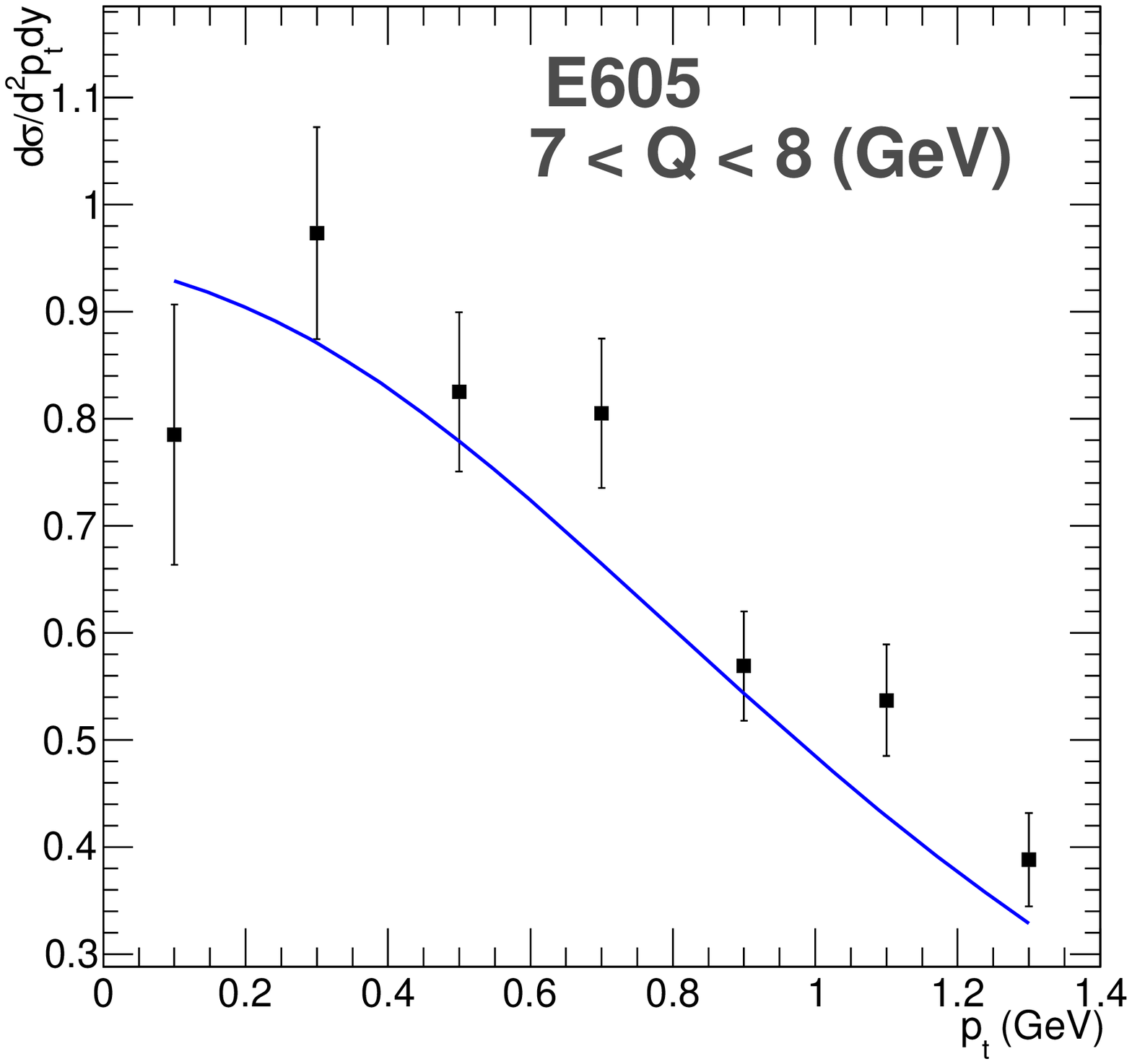}
\includegraphics[width=7cm]{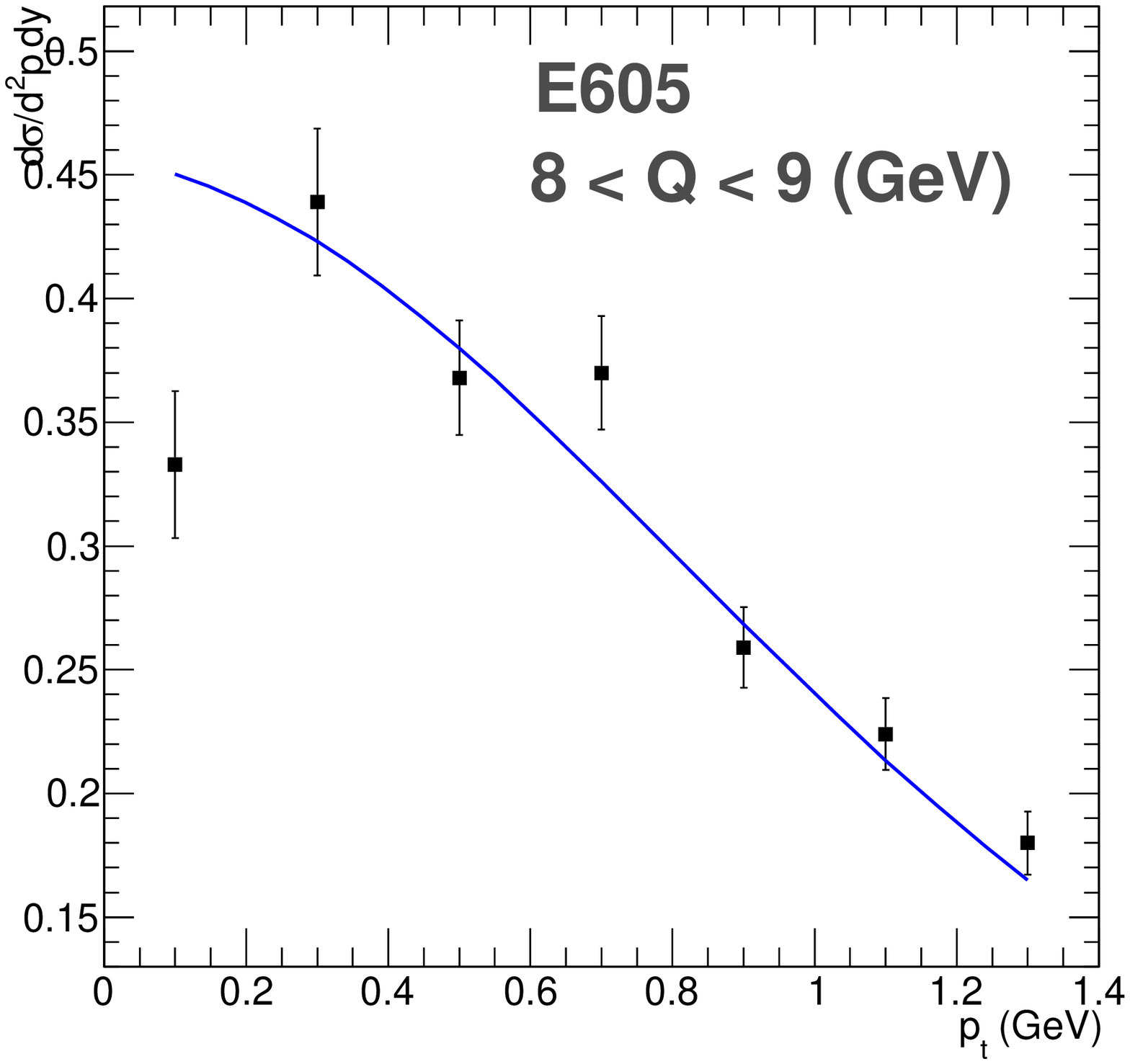}
\includegraphics[width=7cm]{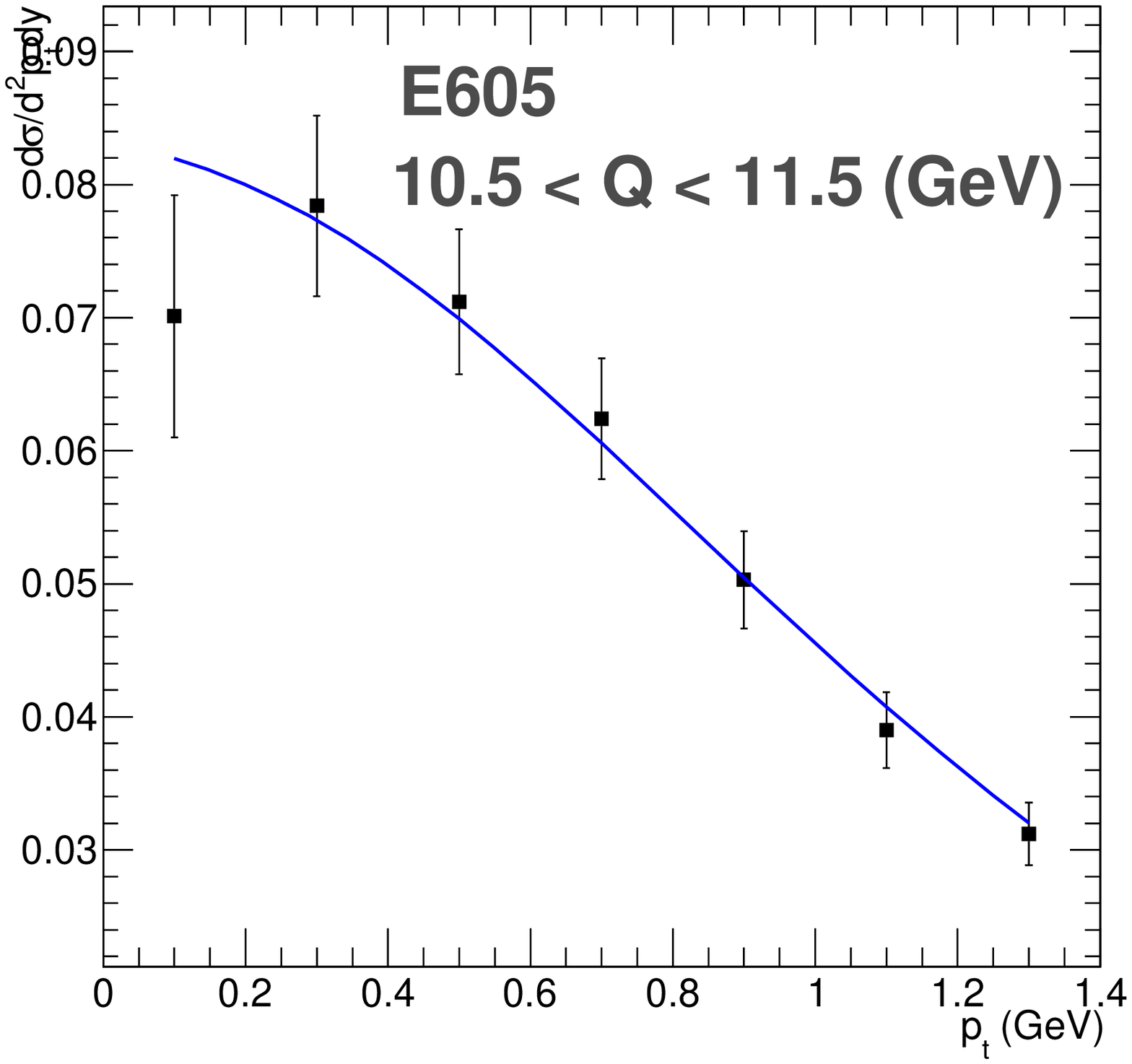}
\includegraphics[width=7cm]{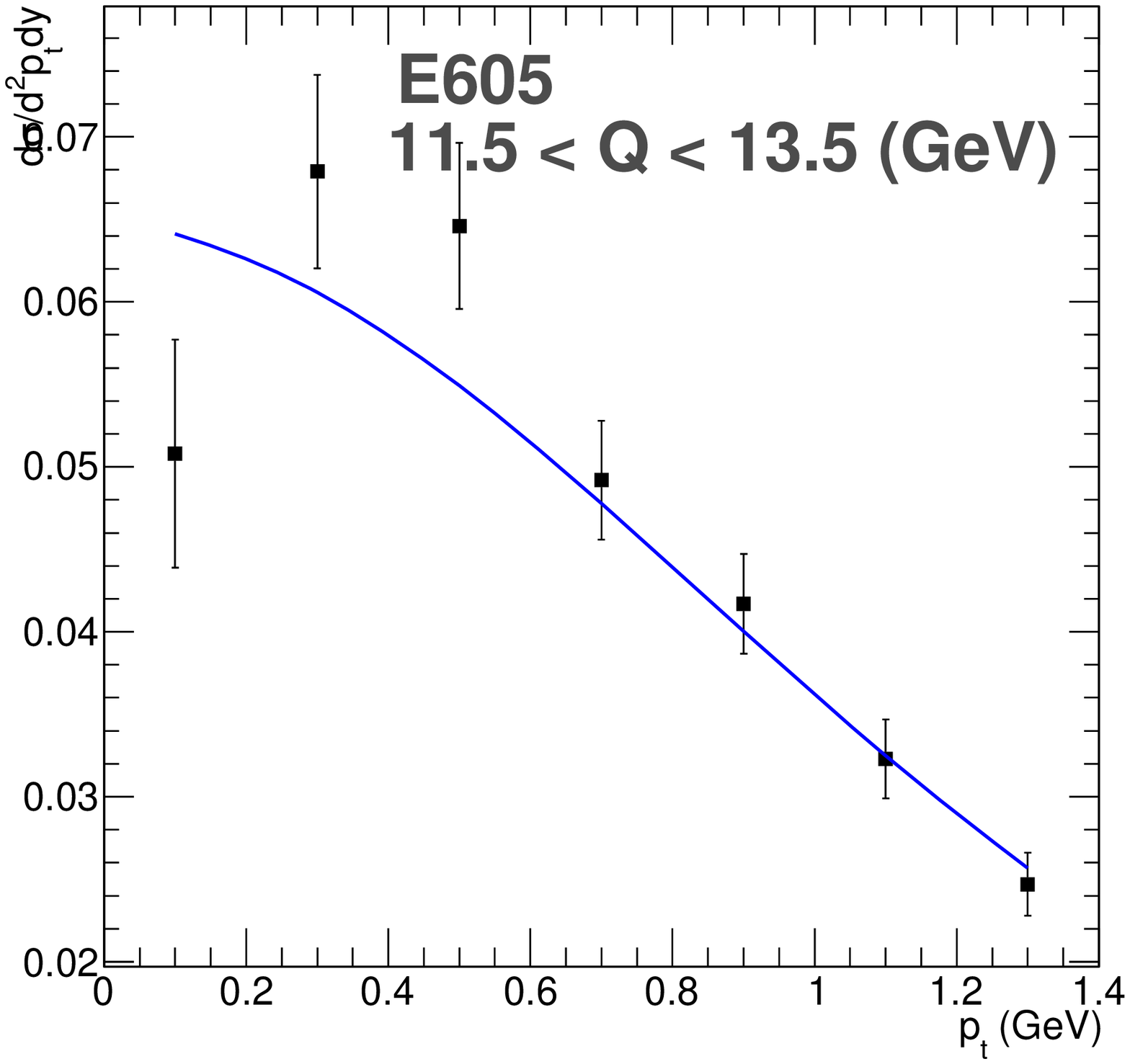}
\includegraphics[width=7cm]{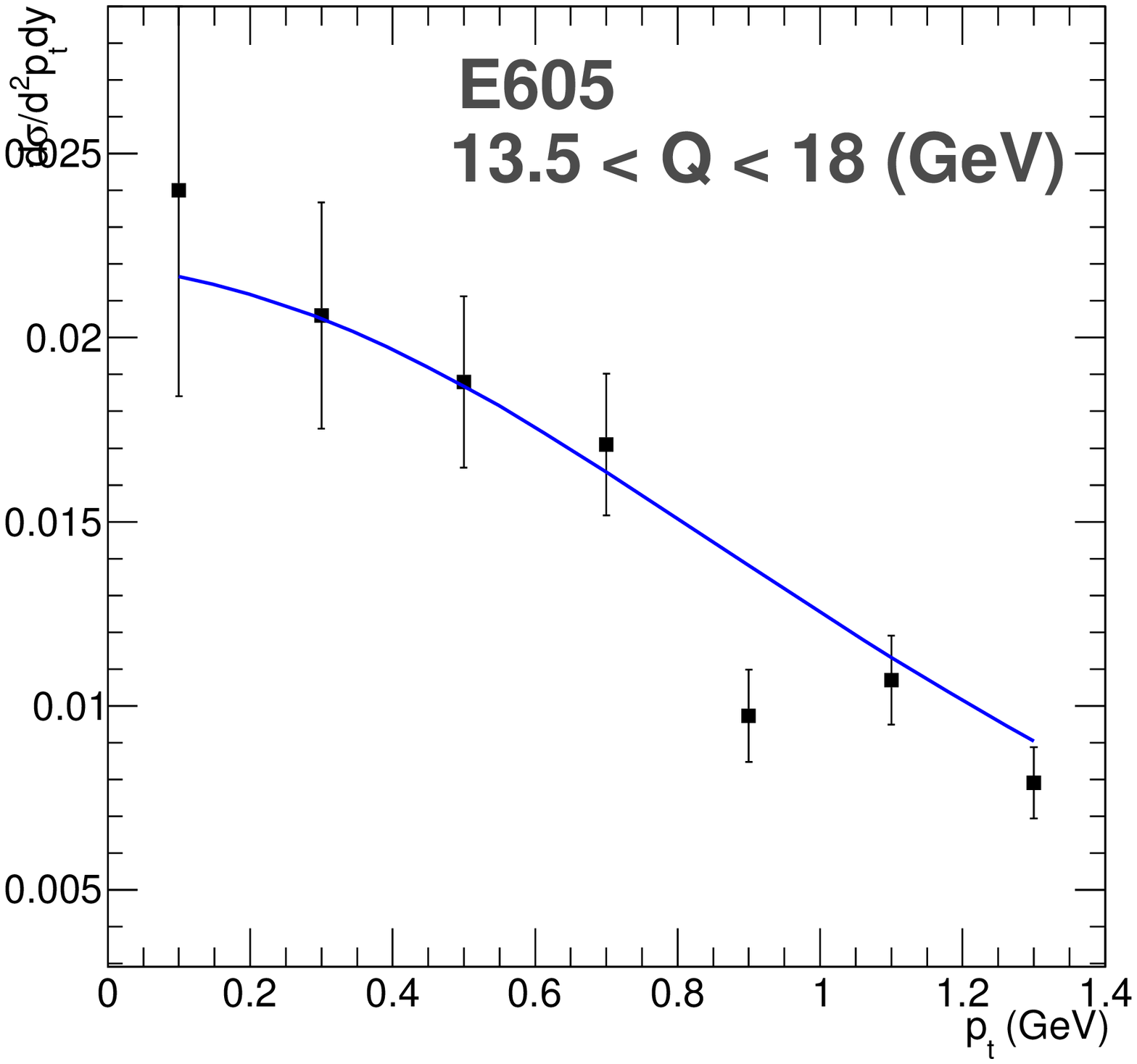}
\caption{Fit to the Drell-Yan data from the E605 Collaboration~\cite{Moreno:1990sf}.}
\label{fig:e605}
\end{figure}

\begin{figure}[tbp]
\centering
\includegraphics[width=7cm]{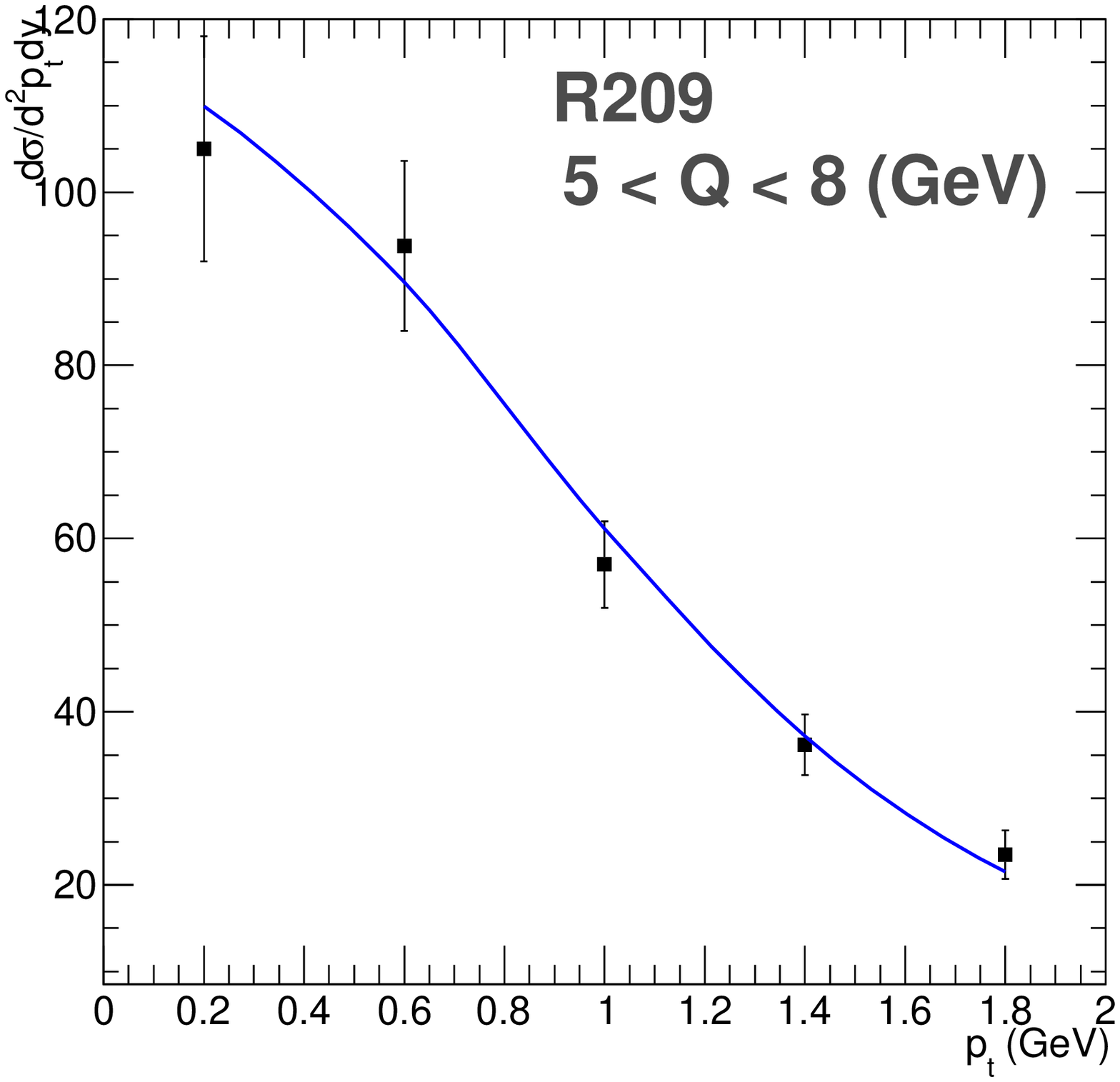}
\includegraphics[width=7cm]{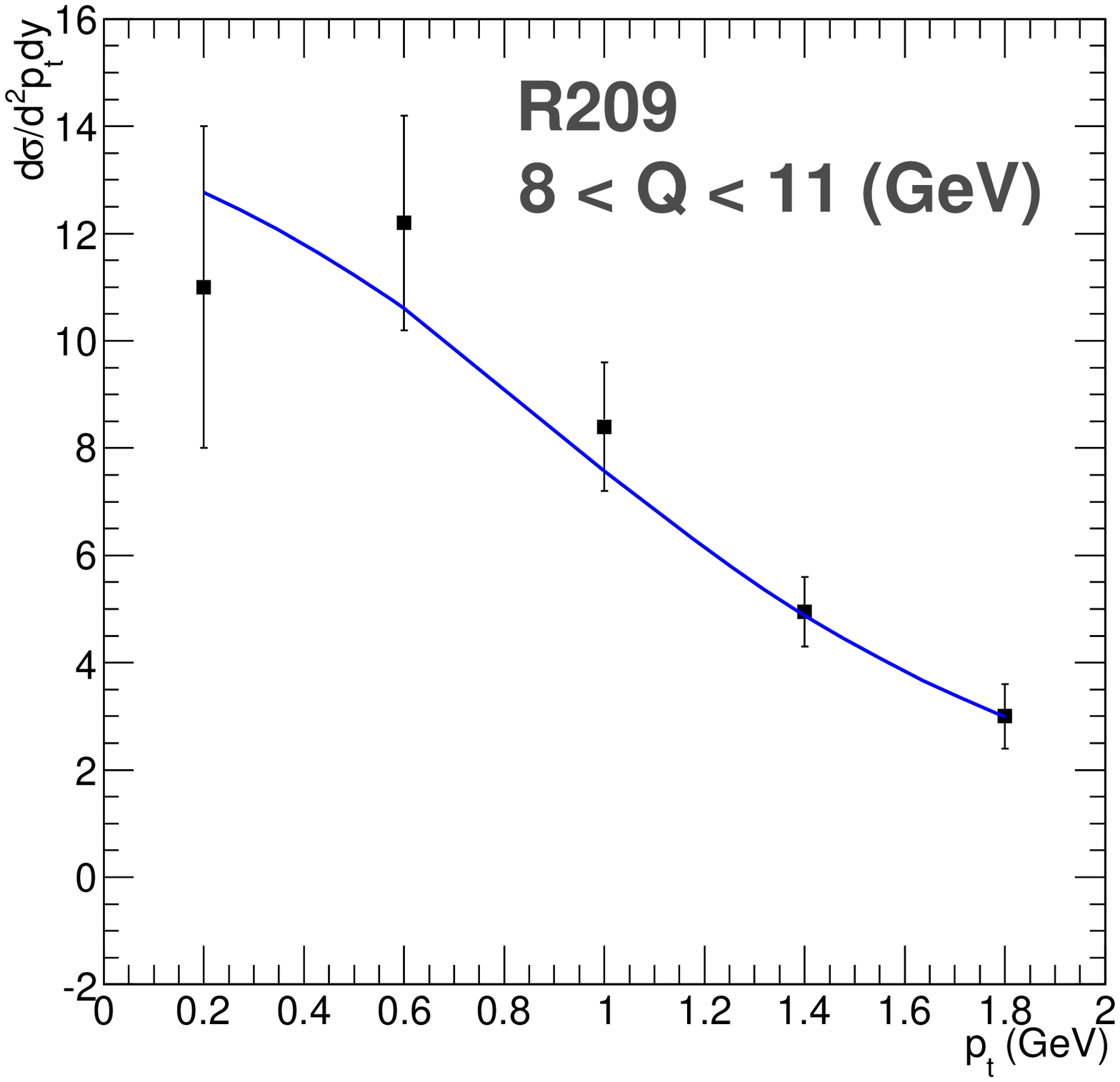}
\caption{Fit to the Drell-Yan data from the R209 Collaboration~\cite{Antreasyan:1981uv}.}
\label{fig:r209}
\end{figure}

\begin{figure}[tbp]
\centering
\includegraphics[width=7cm]{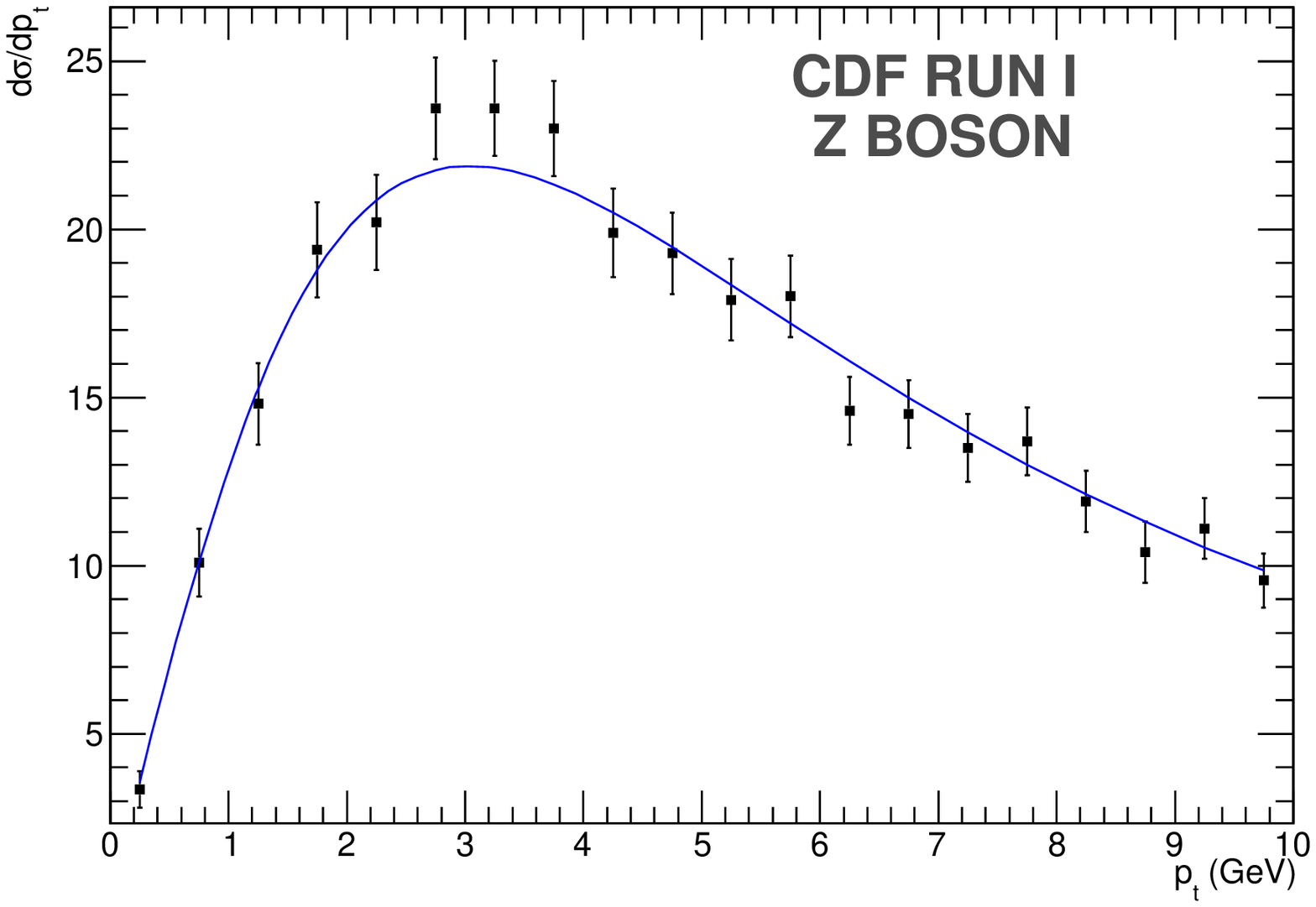}
\includegraphics[width=7cm]{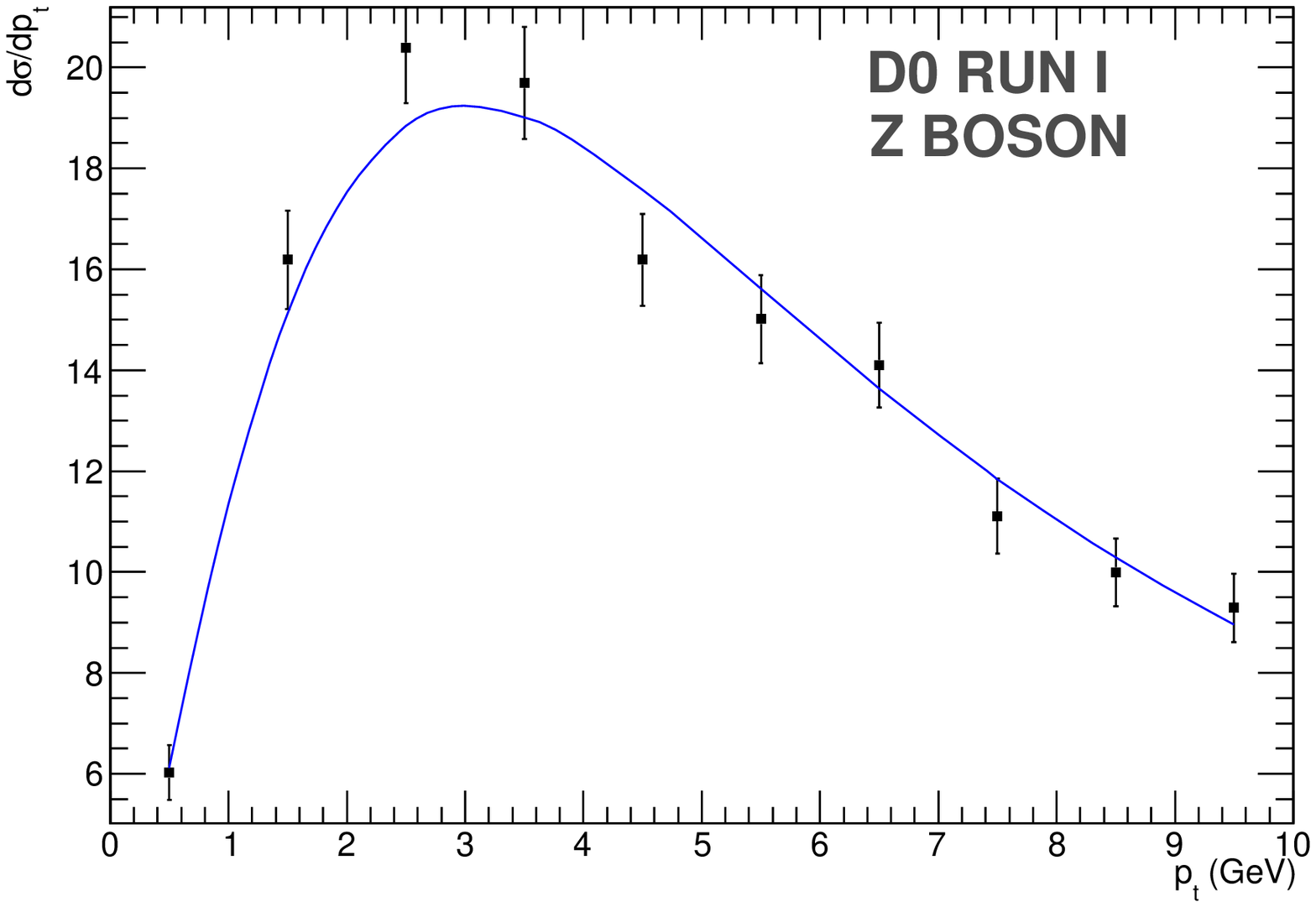}
\caption{Fit to the Tevatron Run I data from the CDF and D0 Collaborations~\cite{Affolder:1999jh,Abbott:1999wk}.
The fits include only the $A^{(1,2)}$, $B^{(1,2)}$, and $C^{(1)}$ contributions. }
\label{fig:tevatron1}
\end{figure}

\begin{figure}[tbp]
\centering
\includegraphics[width=7cm]{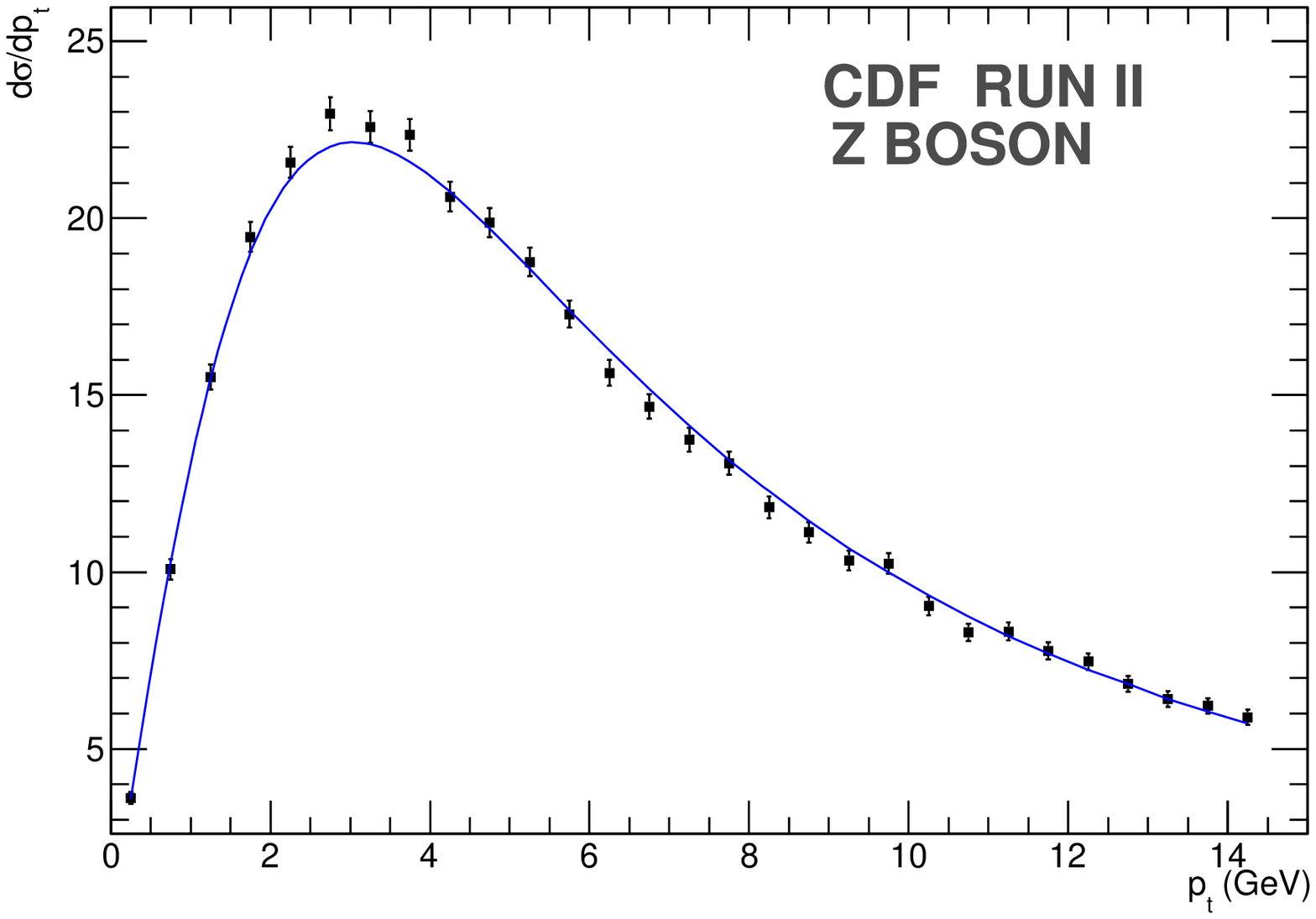}
\includegraphics[width=7cm]{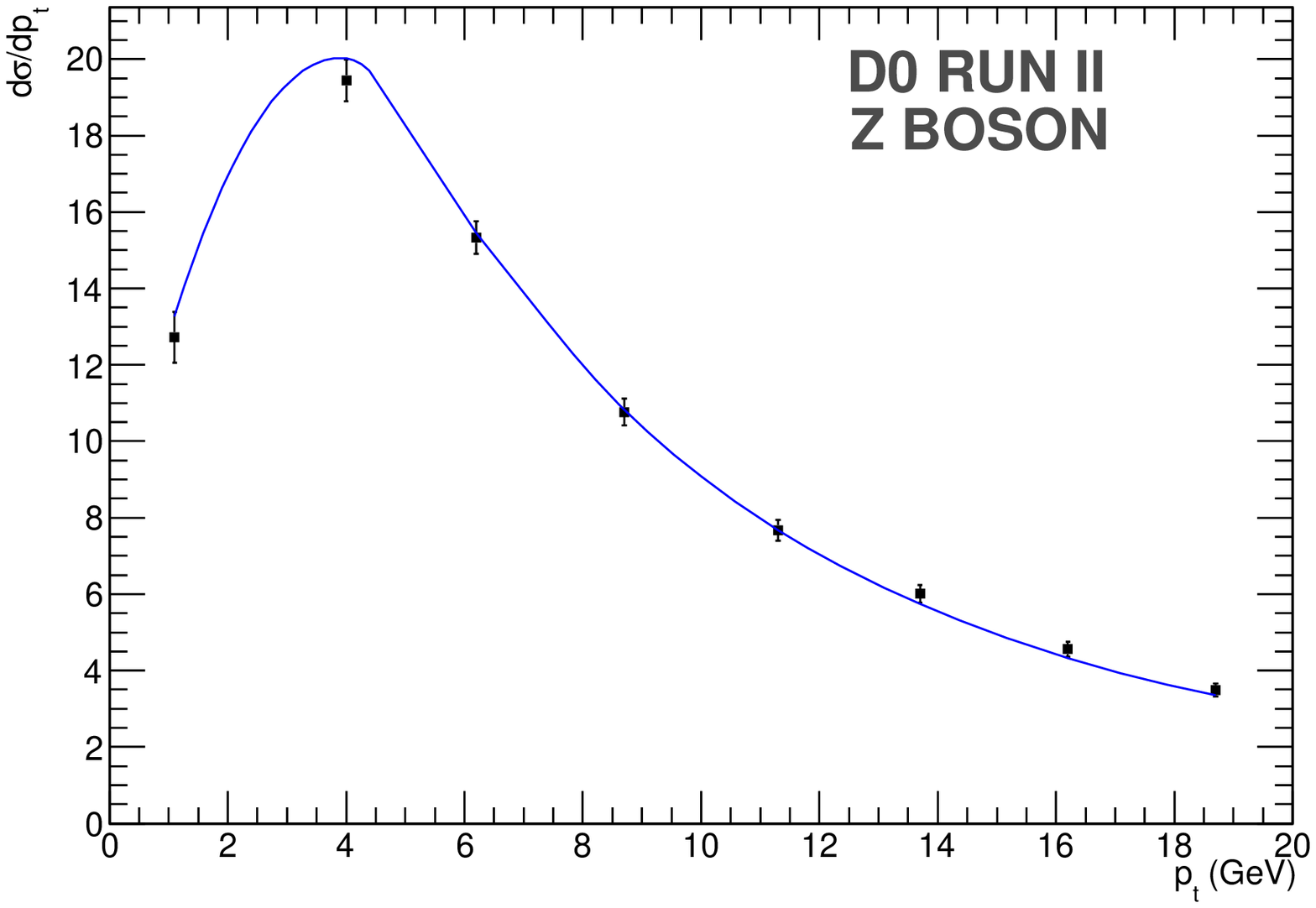}
\caption{Fit to the Tevatron Run II data from the CDF and D0 Collaborations~\cite{Abazov:2007ac,Aaltonen:2012fi}.
}
\label{fig:tevatron2}
\end{figure}

To perform the global analysis of Drell-Yan type processes, 
we include the following data in our fit.
\begin{itemize}
\item Drell-Yan lepton pair production from fixed
target hadronic collisions, including R209, E288 and E605~\cite{Ito:1980ev,Antreasyan:1981uv,Moreno:1990sf}.

\item $Z$ boson production in hadronic collisions from Tevatron
Run I and Run II~\cite{Affolder:1999jh,Abbott:1999wk,Abazov:2007ac,Aaltonen:2012fi}.

\end{itemize}

In total, we include 7 Drell-Yan data sets from 3 fixed target experiments and 4 Tevatron experiments.
Although both CMS and ATLAS have published experimental data
on $Z$ boson production at the LHC, the uncertainties in the present LHC data are
large enough that they do not further constrain the functional form of the above fit.
We will, however, show that 
the theory prediction from our fit can describe the LHC data well.

We would like to emphasize that the high precision data from
$Z$-boson production at the Tevatron Run II~\cite{Aaltonen:2012fi} require
precision calculations of the resummation.
We take 
$g_1$, $g_2$, and $g_3$ as free parameters
in the global fit, and we have chosen $b_{max}=1.5\,{\rm GeV}^{-1}$, 
as proposed in the Konychev-Nadolsky fit~\cite{Konychev:2005iy} 
which takes the exact same form as the BLNY fit.
In the numerical calculations, we adapt the CT10-NLO parton
distribution functions~\cite{Lai:2010vv} at the scale $\mu=b_0/b_*$.
In the resummation calculation, we also take 
into account the running effects of
$\alpha_s$, $\alpha_{em}$, and $N_f$, which are consistent
with the CT10 parameterizations.
These effects are not negligible in the numeric results, and will affect
the fitting parameters.
We have also assigned an additional fitting parameter ($N_{fit}$) for each experiment
to account for the luminosity uncertainties in the experimental measurements. $N_{fit}$ is defined
as a multiplicative factor applied to the theory prediction.

In Figs.~\ref{fig:e288}-\ref{fig:tevatron2}, we show the best fits to the
Drell-Yan data from E288, E605, and R209 Collaborations, and $Z$ boson
production from the CDF and D0 Collaborations at the Tevatron Run I and II.
The results of our fit and the fitted $\chi^2$ values for each experiment 
are listed in Table I.
From these plots, we see that Eq.(\ref{syy}) provides a reasonable fit to all 7 
experiments, with a total of 140 data points,  
with 3 shape parameters $g_{1,2,3}$ and 7 independent normalization factors.
factors. Therefore,
the total number of degrees of freedom in our analysis is 130.

\begin{figure}[tbp]
\centering
\includegraphics[width=7cm]{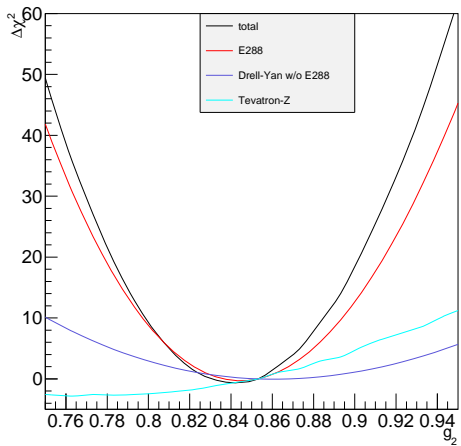}
\caption{$\Delta \chi^2$ distribution scanning $g_2$
parameter in our fit: total and separate contributions
from different experiments: E288, all other Drell-Yan experiments, and
the Tevatron Z-boson experiments, respectively.
}
\label{chi2g2}
\end{figure}

An immediate and important feature from our fit is that the current experimental
data do not provide any useful information on the $x$-dependence of the non-perturbative
form factors, as suggested in Eq.~(\ref{syy}). This is mainly
because the $x$-range covered in these experiments does not reach to
small-$x$ region, in particular for those (low energy)  
fixed target Drell-Yan data.

Among these parameters, the most important one,
relevant to the LHC $W$ and $Z$ boson physics, is $g_2$, which controls
the $Q^2$ dependence in the non-perturbative form factors. 
To obtain the error in the determination of $g_2$ value in our fit, 
we scan $g_2$ around its best fit value and show the variation 
in the total chi-square from the best fit, denoted by $\Delta \chi^2$, 
in Fig.~\ref{chi2g2}.
The $g_2$ error is estimated at the 68\% confidence level (C.L.) by 
taking $\Delta \chi^2$ around 7.3, for 130 degrees of freedom in the $\chi^2$ distribution.
Hence, the $g_2$ value in our fit is 
\begin{equation}
g_2=0.84^{+0.040}_{-0.035}~~(\rm at~68\% ~C.L.) \ .
\end{equation}
In order to demonstrate the sensitivities of 
of various experiments on the determination of the $g_2$ value, 
we further plot the $\Delta\chi^2$ distributions as functions of $g_2$ from
each data set.
From this figure, we can clearly see that the most strong constraints come from the precision
Drell-Yan data at fixed target experiments, i.e., the E288 experiment.
Although the Tevatron data on the $Z$-boson production is the most precise Drell-Yan type
data in hadronic collisions, they do not pose a strong constraint on the non-perturbative
form factor $g_2$. This is due to the fact that the energy at the Tevatron is much higher, and
therefore is dominated by the perturbative Sudakov factor instead of the non-perturbative Sudakov
for $W$ and $Z$ boson production at higher energies.
This also will hold true for the LHC since it is even higher energy than the Tevatron.
 Similar observation
has also been obtained in Ref.~\cite{Qiu:2000ga} with different prescription
of implementing the non-perturbative form factors in the CSS resummation 
formalism. 

\begin{figure}[tbp]
\centering
\includegraphics[width=8cm]{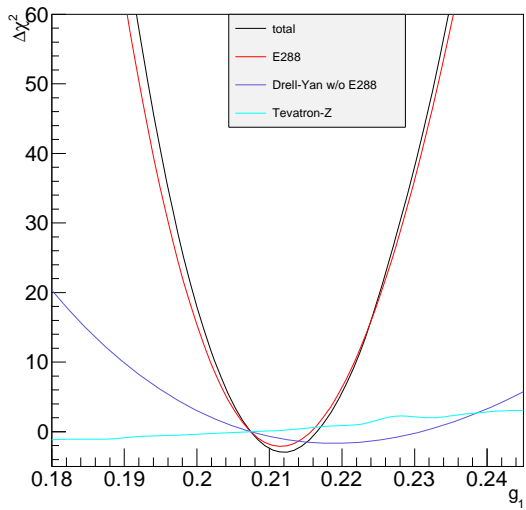}
\caption{Same as Fig.~\ref{chi2g2} for $g_1$.
}
\label{chi2g1}
\end{figure}

As we mentioned above, the $g_2$ term in the non-perturbative form factor
scales as as $b^2\ln(Q)$ at small $b$, 
because $\ln(b/b_{\star}) \sim b^2/(2 b_{max}^2) $ for $b \ll b_{max}$.
By using the above parameter, we find that the small-$b$
behavior of our fit can be written as $0.187b^2\ln(Q)$ which is in the similar
range of the fit found in Ref.~\cite{Konychev:2005iy} with the same 
choice of $b_{max}=1.5\,{\rm GeV}^{-1}$.
It is interesting to note that the $g_2$ value 
can also be estimated from fixed order calculations,
from which we find that $g_2\approx 4C_F\alpha_s/\pi$~\cite{Collins:1985xx}. 
Therefore, the fitted $g_2$ value implies 
$\alpha(\mu) \sim 0.49$, which suggests that the relevant 
nonperturbative physics effect sets in around $\mu \sim 1$ GeV, 
the same order as $b_{max}$ used in this analysis.

Similarly, we examine in Fig.~\ref{chi2g1} the sensitivity
of various experiments on the determination of the $g_1$ value. 
The major contribution to the $\Delta \chi^2$ again comes
from fixed target Drell-Yan experiments. 
Moreover, the $g_1$ value in our fit is found to be 
\begin{equation}
g_1=0.212^{+0.006}_{-0.007}~~(\rm at~68\% ~C.L.) \ .
\end{equation}

\begin{figure}[tbp]
\centering
\includegraphics[width=9cm]{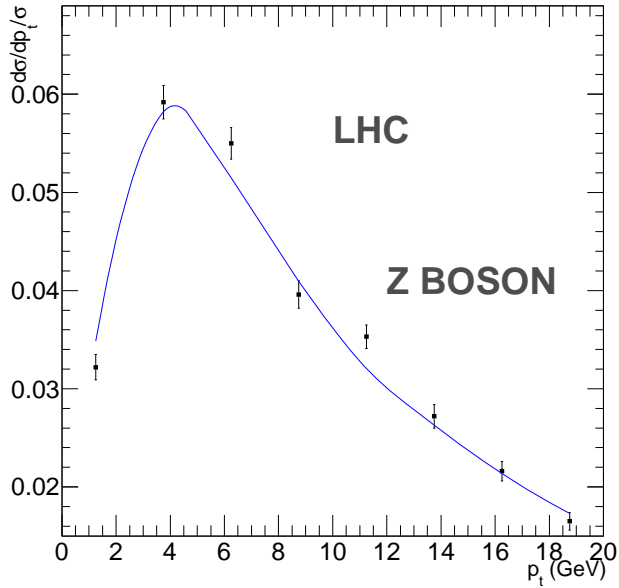}
\caption{Compare the resummation prediction for $Z$ boson production at the
LHC~\cite{Ito:1980ev}. These data are not included in our fit.}
\label{fig:lhc}
\end{figure}

Recently, both CMS and ATLAS Collaborations have published their data
on $Z$ boson production at the LHC. We compare our predictions to the ATLAS data~\cite{Aad:2011gj}
in Fig.~\ref{fig:lhc}. From this figure, we can see that our fit can describe the
LHC data well.

Before we check the consistency between the above fitting results with the SIDIS
data from HERMES/COMPASS, we would like to emphasize that the above parameters
are fitted only with the Drell-Yan type data. From the comparison to
the experimental data, we can see that the new form is equally good as compared
to the original BLNY parameterization.
We will discuss more about this comparison in the Conclusion section.

\section{Semi-inclusive DIS with the New Parameterizations}

The universality of the parton distribution is a powerful prediction from
QCD factorization. According to the TMD factorization,
 we will expect the universality of the TMD parton distributions
between SIDIS and Drell-Yan processes as well. Therefore, the non-perturbative
functions determined for the TMD parton distributions from the Drell-Yan type of
processes shall apply to that in the SIDIS. Of course, the transverse momentum
distribution of hadron production in DIS processes also depends on the final state
TMD fragmentation functions, which 
need to be determined by fitting to existing experimental data.
Following the universality arguments,
we assume the following parameterizations for the non-perturbative form factors for
SIDIS process, in contrast to  Eq.~(\ref{syy}) for Drell-Yan process, 
\begin{eqnarray}
S_{NP}^{(DIS)}&=&
{g_1 \over 2} b^2
+ g_2 \ln\left(b/b_*\right)\ln(Q/Q_0)
+ g_3 b^2 (x_0/x_B)^\lambda
+ {g_h \over z_h^2} b^2 \ . 
\label{syy2}
\end{eqnarray}
In the above parameterization, $g_1$ and $g_2$ have been determined
from the experimental data of Drell-Yan lepton pair production.
The factor of $1/2$ in front of the $g_1$ term is due the fact 
that there is only one incoming hadron in the SIDIS process, 
while there are two incoming hadrons in the Drell-Yan process.
Although there has been evidence
from recent studies~\cite{Signori:2013mda,Aidala:2014hva}
that $g_h$ could be different for the so-called favored and dis-favored fragmentation
functions, we still take them to be the same in this study for simplicity. 
When more precise data become available, we may need to perform a global 
analysis with two separate $g_h$ parameters. 

In principle, we can fit $g_1$, $g_2$, and $g_h$ together to both Drell-Yan
and SIDIS data. However, the SIDIS data from HERMES and COMPASS
mainly focus in the relative low $Q^2$ range. Because of that, the theoretical 
uncertainty of the CSS prediction is not well under controlled, particularly, 
from the $Y$-term contribution which will be discussed in the following subsection.
There have been several successful phenomenological studies to describe 
the experimental data from HERMES
and COMPASS experiments, using the leading order TMD 
formalism ~\cite{Signori:2013mda,Anselmino:2013lza}.
The goal of this paper is to check if we can apply the non-perturbative form factors
determined in the Drell-Yan process to the SIDIS processes.
As shown in Ref.~\cite{Sun:2013dya}, we can not do that with
the original BLNY or KN fit, where it was found that the extrapolation of these
fits to the kinematic region of HERMES and COMPASS is in conflict with the
experimental data. We will show, however, the SIYY form will be able to extend to
SIDIS experiments from HERMES and COMPASS Collaborations.

\begin{figure}[tbp]
\centering
\includegraphics[width=6cm]{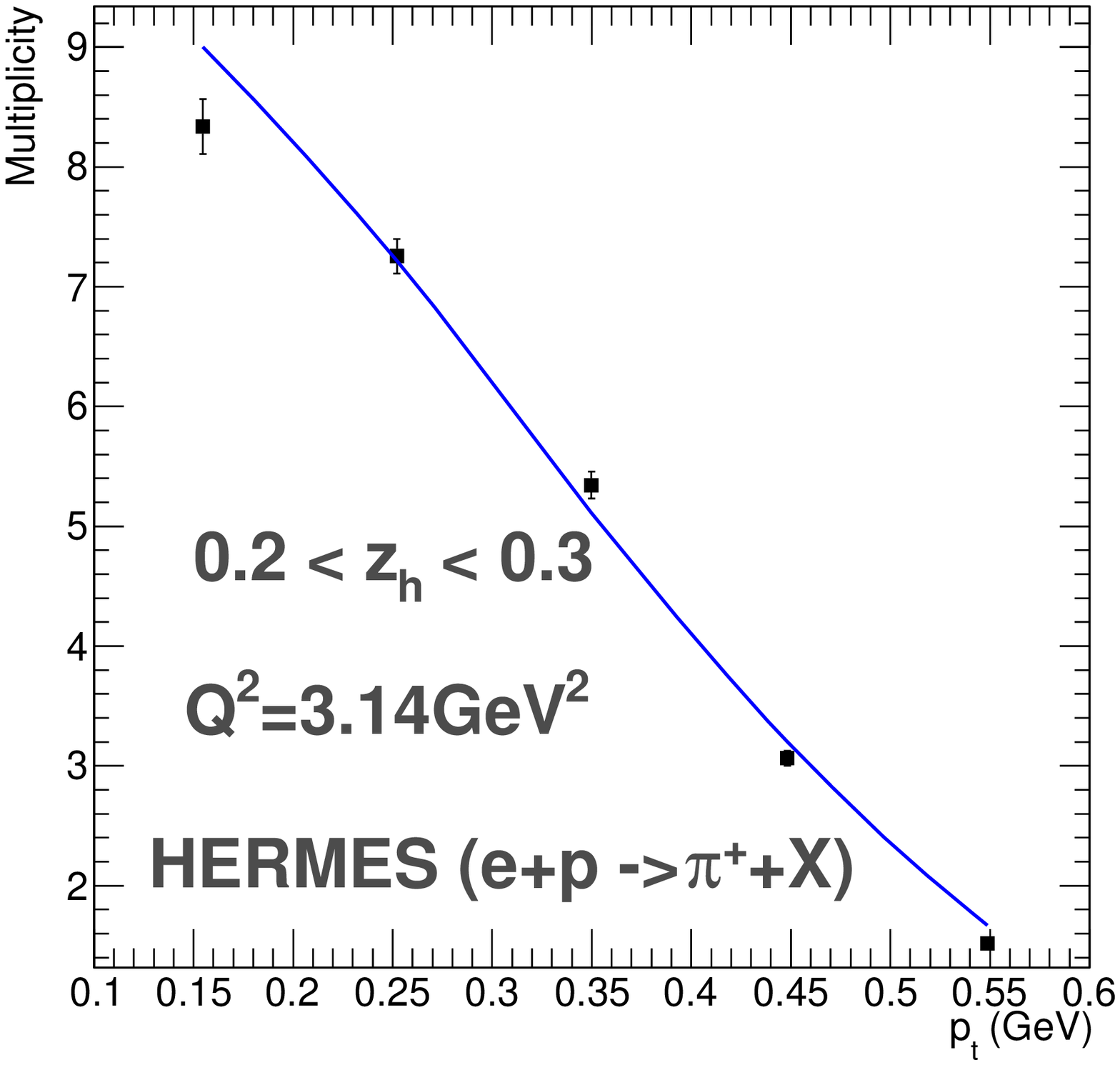}
\includegraphics[width=6cm]{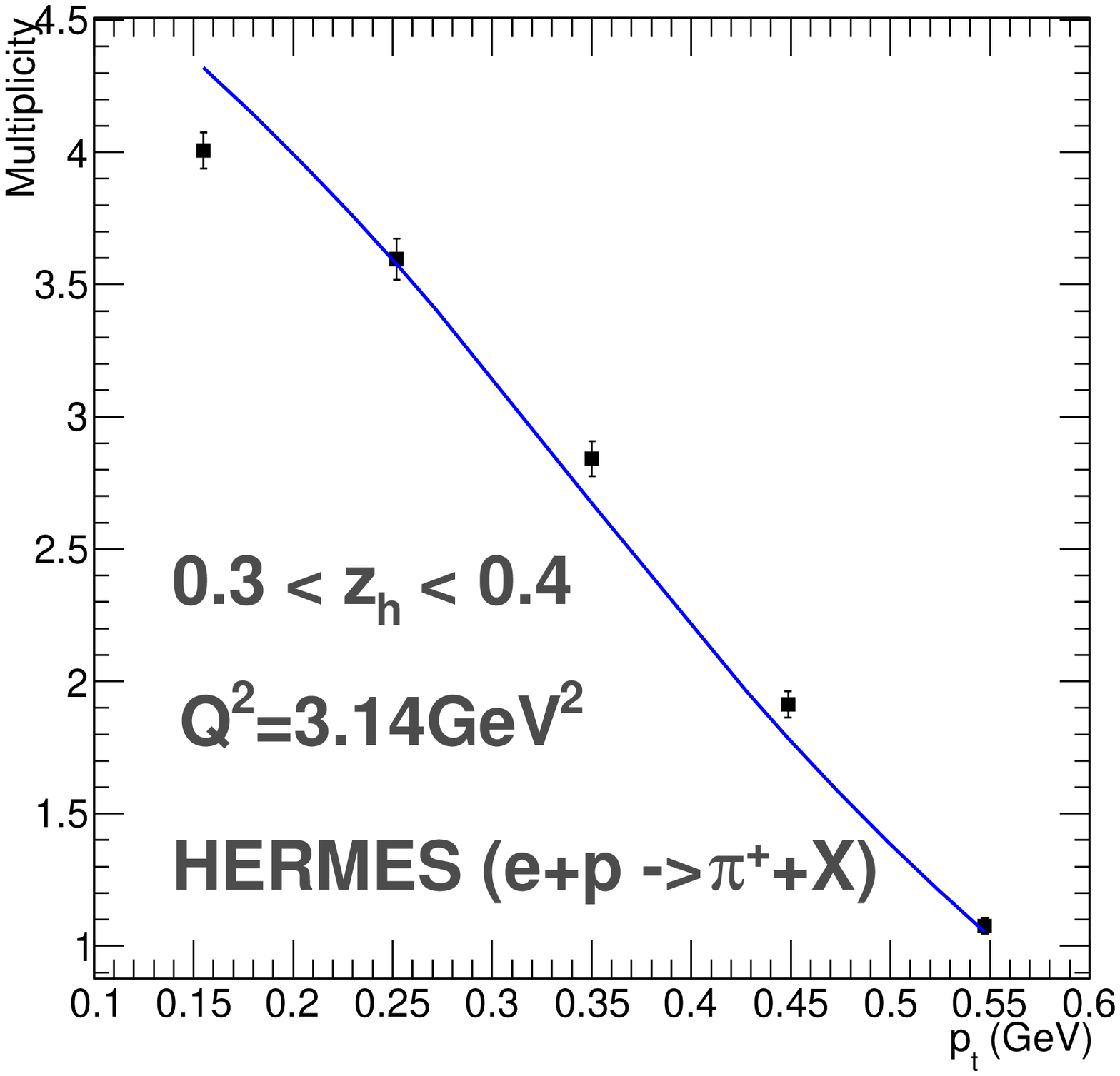}
\includegraphics[width=6cm]{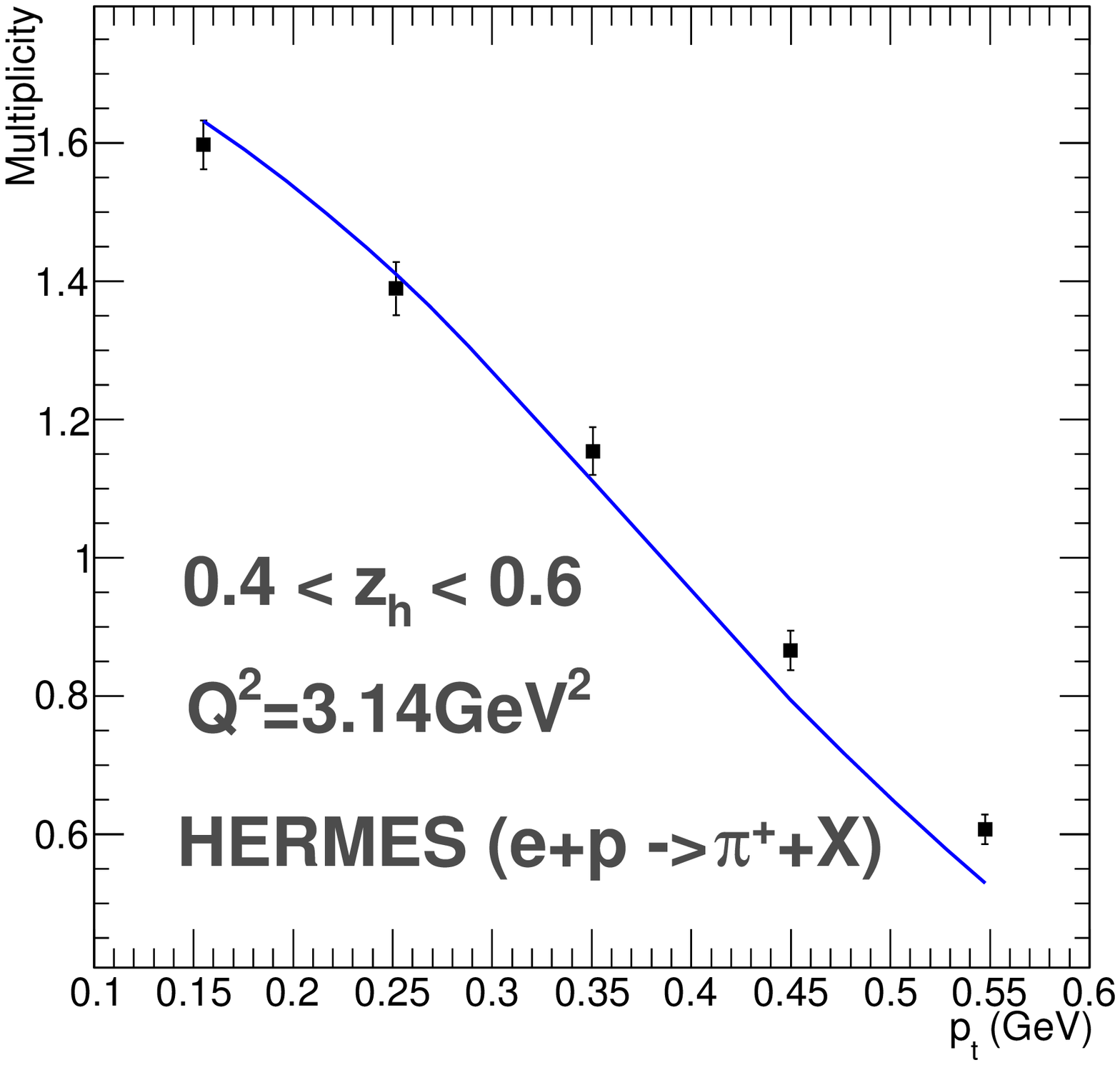}
\includegraphics[width=6cm]{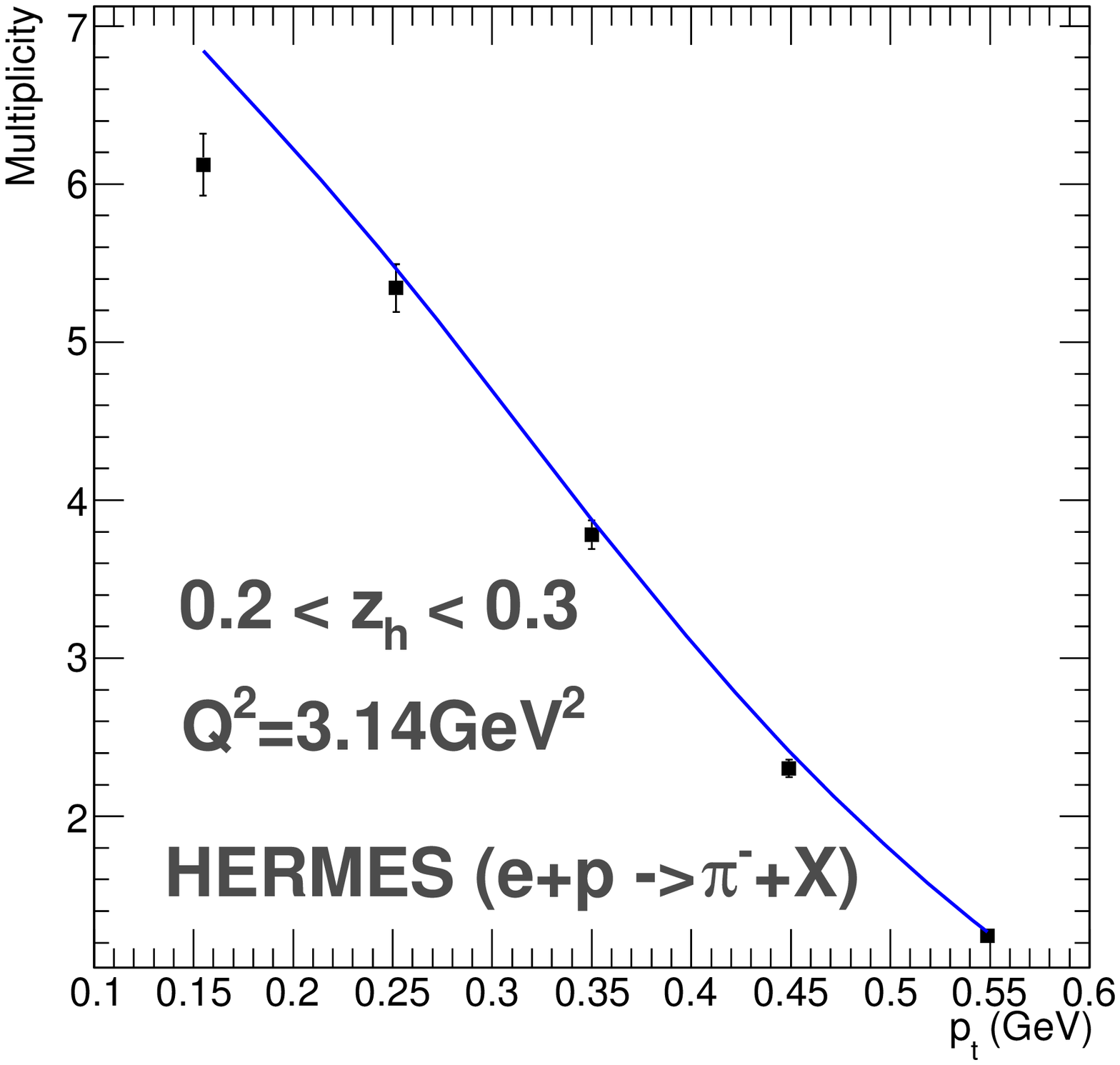}
\includegraphics[width=6cm]{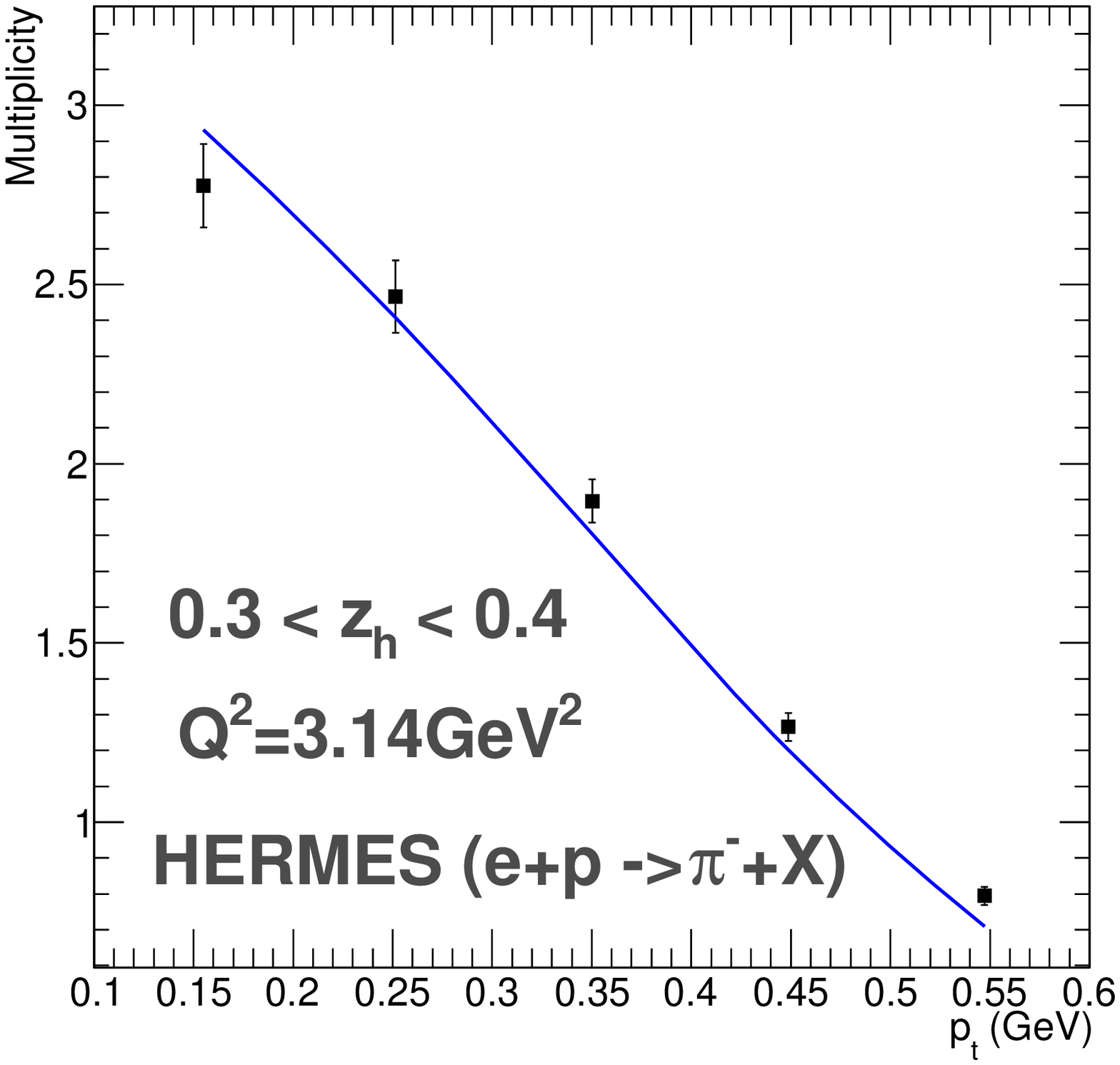}
\includegraphics[width=6cm]{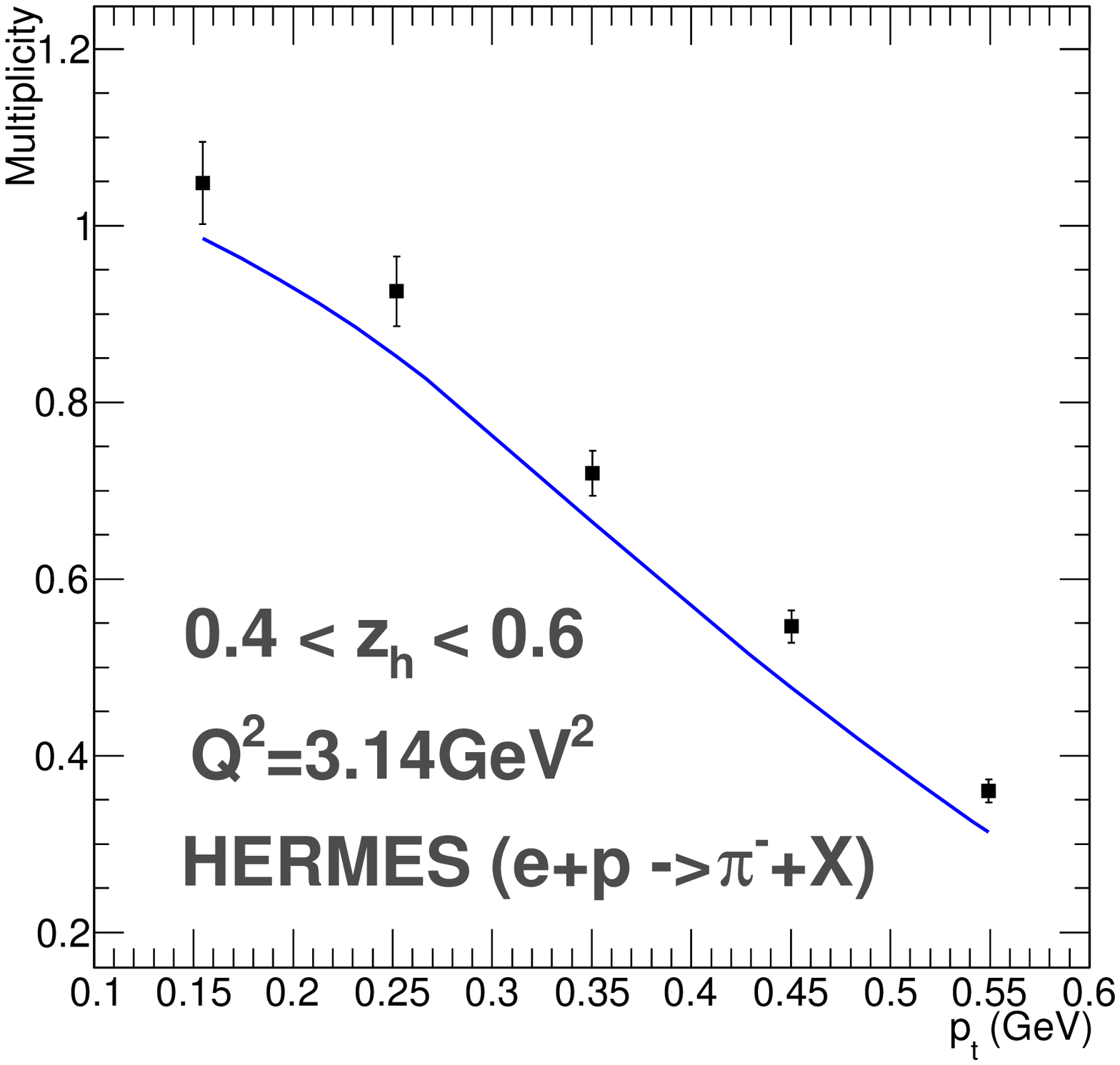}
\caption{Multiplicity distribution as function of transverse
momentum in semi-inclusive hadron production in deep inelastic
scattering compared to the experimental data from HERMES
Collaboration at $Q^2=3.14$GeV$^2$.}
\label{fig:hermes}
\end{figure}

Therefore, in the following, we will take the parameters ($g_{1,2}$) fitted to the
Drell-Yan data to compare to the SIDIS to check if they are consistent with
the SIDIS data.
In Fig.~\ref{fig:hermes}, we show the comparisons between the theory
predictions with $g_h=0.042$ and the SIDIS data from HERMES,
with total $\chi^2$ around 180. This parameter is consistent with previous
analysis when leading order TMD formalism is considered~\cite{Signori:2013mda,Anselmino:2013lza}.
It is also consistent with the TMD formalism with truncated evolution effects
in Ref.~\cite{Sun:2013dya}.
The differential cross section for SIDIS process depends on the hadron
fragmentation functions, for which we adopt the parameterization
from the new DSS fit~\cite{deFlorian:2007aj,deFlorian:2014xna}. We include
a normalization factor about 2.0 in the calculation of the multiplicity distributions
shown in Fig.~\ref{fig:hermes}, which accounts for theoretical uncertainties
from higher order corrections for both differential and inclusive cross sections~\footnote{
Compared to the leading order TMD fit of Ref.~\cite{Anselmino:2013lza} where there is
no normalization factor, the $C^{(1)}$ coefficient is large and negative in the CSS
resummation application to the SIDIS. Phenomenologically, that is the reason we have
to include a factor of 2 in the comparison to the SIDIS data. This could be improved
if the differential cross section (instead of multiplicity distributions) can be measured in the future.}.
Here, the $Y$-term contribution is not included, which will be discussed 
in the following subsection.

Figs.~\ref{fig:e288}-\ref{fig:hermes} clearly illustrate that we have obtained a universal non-perturbative
TMD function which can be used to describe both Drell-Yan lepton
pair production and semi-inclusive hadron production in DIS
processes in the CSS resummation framework. We also want to point
out that the new functional form for the non-perturbative
function is crucial to achieve this conclusion 
as given in Eqs.~(\ref{syy}) and (\ref{syy2}).

\subsection{Issue with the $Y$ Term in SIDIS for HERMES and COMPASS}

\begin{figure}[tbp]
\centering
\includegraphics[width=8cm]{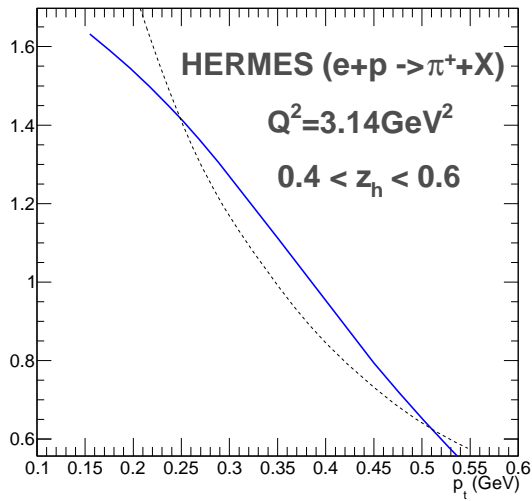}
\caption{$Y$-term contribution (dashed  curve) to the multiplicity distribution
as a function of transverse momentum, compared to
 the leading power transverse momentum dependent result (solid curve), 
 for the experimental data from HERMES Collaboration at $Q^2=3.14\,{\rm GeV}^2$.
 }
\label{fig:hermesy}
\end{figure}

In Fig.~\ref{fig:hermes}, we have neglected the contribution from the $Y$-term.
This may be a strong approximation for HERMES
and COMPASS experiments because their data are typically in the relative low
$Q^2$ range. Indeed, we find that the numeric contributions from
$Y$-term are important for both HERMES and COMPASS
experiments. One example is shown in Fig.~\ref{fig:hermesy} for
$z_h=0.4$-$0.6$. The dashed curve represents the
$Y$-term contribution, whereas the solid curve represents the resummation
prediction without including the $Y$-term.
It appears that adding the $Y$-term contribution will worsen the
agreement between the theory prediction and the experimental data.
Numerically, the $Y$-term contribution is at the
same order of magnitude as the leading power contribution 
in the TMD resummaiton formalism, which is formally defined 
as the resummation calculation without including the $Y$-term contribution.
At a smaller $z_h$ value, the  $Y$-term contribution becomes even more 
important as compared to the the leading power TMD contribution.

This is an important observation, and raises a concern on the
interpretation of the 
existing SIDIS data whose relevant energy scale is low, on the order of a few GeV.
Theoretically, it indicates
that higher order corrections in $Y$-term are important and may have
to be taken into account to understand the experimental data.
The dashed curves in Fig.~\ref{fig:hermesy} only include $Y^{(1)}$ contribution. $Y^{(2)}$ for
SIDIS has not yet been calculated in the literature. We hope to carry out
this computation and come back to this issue
in the near future.
This may also indicate that we need to take into
account higher power corrections for SIDIS processes
in the relative low $Q^2$ range. In this context, it means that certain
terms in the $Y$-term may come from higher power correction in the
TMD factorization, which could result in different resummation results.
This is similar to what has been discussed in Ref.~\cite{bbdm} for higher-twist
contributions to the SIDIS, where $\cos \phi$ and $\cos 2\phi$ azimuthal
asymmetries in SIDIS processes come from higher-twist effects in the TMD
framework. However, the factorization for higher-twist contribution in the
TMD framework is not fully understood at the present.

\begin{figure}[tbp]
\centering
\includegraphics[width=6cm]{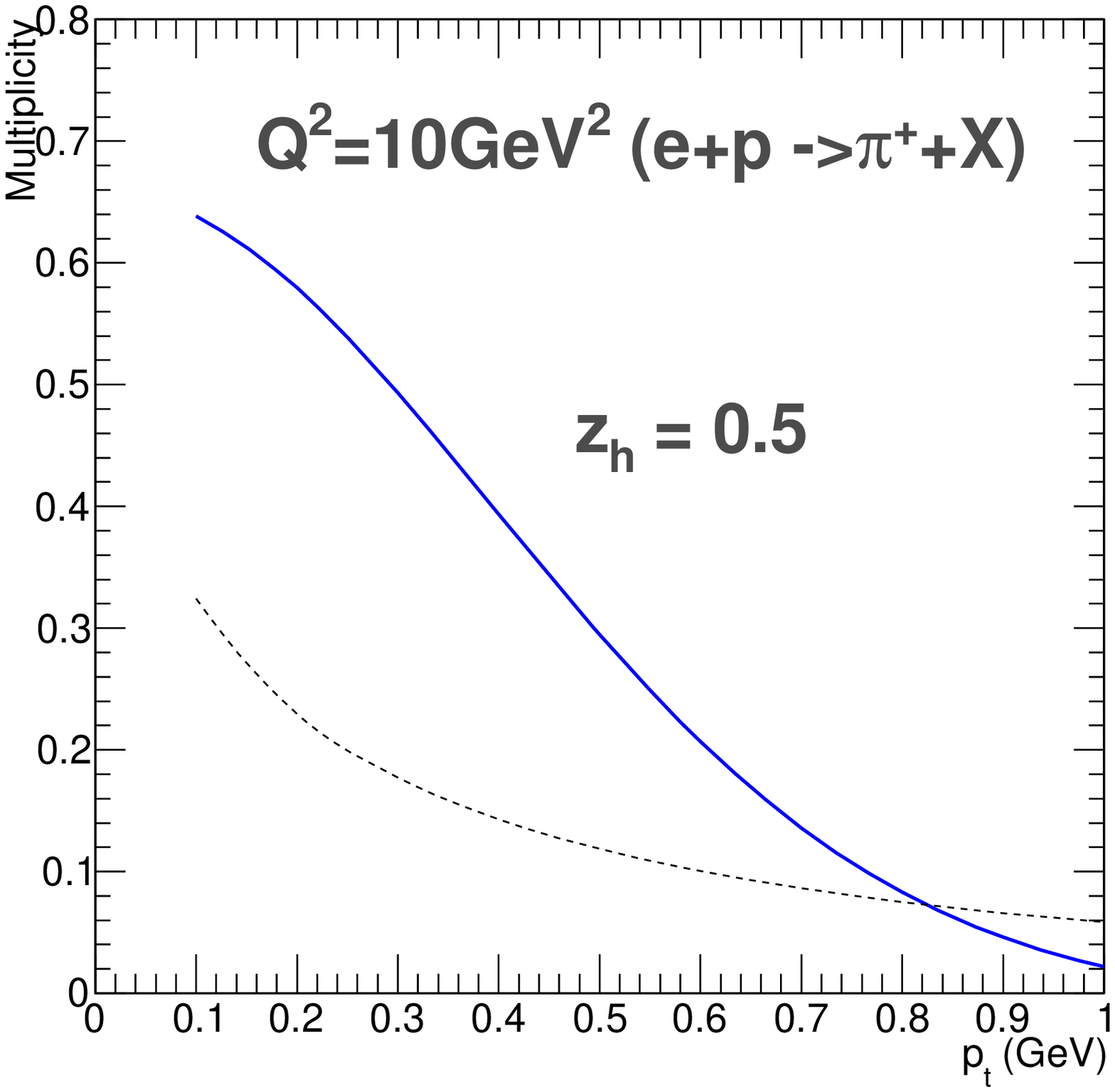}
\includegraphics[width=6cm]{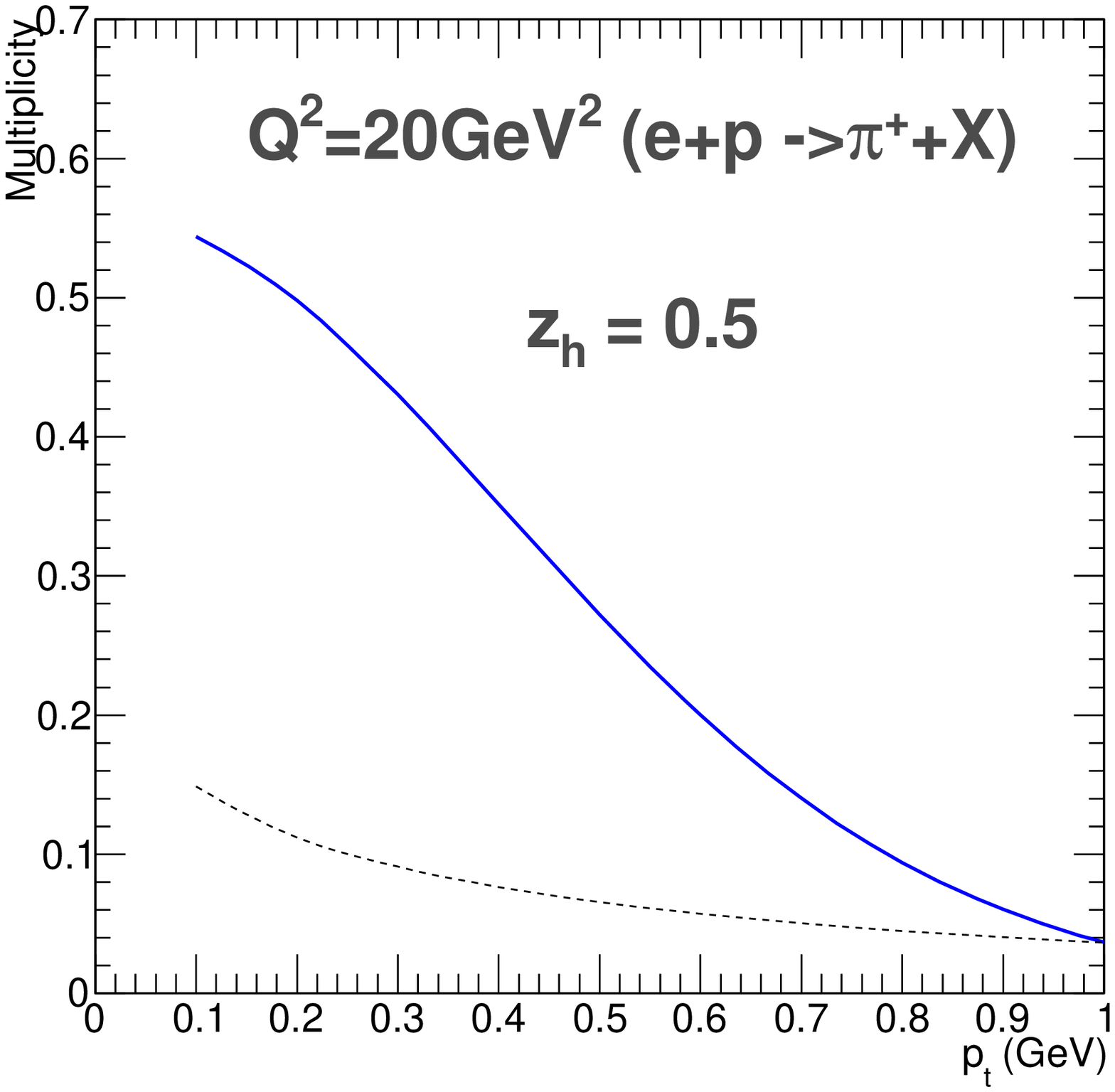}
\caption{Comparison between the leading power TMD calculations (solid curves) and the
$Y$-term contributions (dashed curves) for $Q^2=10\,{\rm GeV}^2$ (left) 
and $Q^2=20\,{\rm GeV}^2$ (right)
for typical values of $x_B=0.1$ and $z_h=0.5$.}
\label{fig:hermesyp}
\end{figure}

On the other hand, the consistency between the leading power TMD results and the
experimental data from HERMES and COMPASS collaborations, cf. Fig.~\ref{fig:hermes}, 
supports the application of the TMD factorization in the relative low $Q^2$
range of these two experiments. To further test the TMD resummation formalism 
in the SIDIS experiments, we need more data with large $Q^2$ values, where 
the $Y$ -term contributions will become much less important. 
In Fig.~\ref{fig:hermesyp}, we show some numeric results for $Q^2=10$, $20$ GeV${}^2$.
In particular, for $Q^2=20\,{\rm GeV}^2$, its contribution is negligible for all $p_\perp$
range of interests. Higher $Q^2$ range is particularly one of the important focuses
for the SIDIS measurements in the planned electron-ion collider~\cite{Boer:2011fh},
where the above assumptions can be well tested.

\section{Discussion and Conclusion}

In this paper, we have re-analyzed the transverse momentum
distribution of the Drell-Yan type of lepton pair production processes
in hadronic collisions in the framework of CSS resummation formalism.
Our goal is to find a new form for the non-perturbative function which can
be used to simultaneously describe the semi-inclusive hadron production
in DIS processes (such as from HERMES and COMPASS Collaborations)
and all the Drell-Yan type processes (such as $W$, $Z$ and low energy
Drell-Yan pair productions).
In Secs. II and III, we argue for a new parametrization form, Eq.~(\ref{syy}),
for describing Drell-Yan processes.
For clarity, we recap our findings, and name it as the SIYY-1 form,
as follows.
\begin{equation}
S^{\rm SIYY-1}_{NP}=g_1b^2+g_2\ln\left(b/b_*\right)\ln\left({Q}/{Q_0}\right)
+g_3b^2\left((x_0/x_1)^\lambda+(x_0/x_2)^\lambda\right)
\ ,\label{syy-1}
\end{equation}
where we adopted the $b_*$ description, cf. Eq.~(2), with
$b_{max}=1.5\,{\rm GeV}^{-1}$,
and
have fixed $Q_0=1.55$\,GeV, $x_0=0.01$ and $\lambda=0.2$ in a
global analysis of the low energy Drell-Yan data from E288, E605, R209,
and $Z$ boson data from CDF and D0 at the Tevatron (in both Run I and II).
In total, we have included 140 data points, fitted
with 3 shape parameters ($g_1, g_2, g_3$) and 7 normalization parameters.
The chi-square per degree of freedom is about 1.3, cf. Table I.
We found that at the 68\%~C.L.,
\begin{eqnarray}
g_1&=&0.212^{+0.006}_{-0.007}\ , \nonumber \\
g_2&=&0.84^{+0.040}_{-0.035}\ , \nonumber \\
g_3&=& 0.0 \ .
\end{eqnarray}
The detailed comparison of the fit to the
experimental data can be found in Figs. 1 to 7.
Using the result of the fit, we showed in Fig. 8 that
the LHC data can also be well described by the SIYY-1 fit.

After obtaining the satisfactory fit to the Drell-Yan type data, we
proposed to add an additional term to the SIYY-1 form with the
$z_h$ dependence for describing the transverse momentum
distribution of the semi-inclusive hadron production
in DIS processes, cf. Sec. IV.
We shall name that as the SIYY-2 form, which is
\begin{equation}
S^{\rm SIYY-2}_{NP}={g_1 \over 2} b^2+g_2\ln\left(b/b_*\right)\ln\left({Q}/{Q_0}\right)
+g_3b^2(x_0/x_B)^\lambda+{g_h \over z_h^2} b^2
\ ,\label{syy-2}
\end{equation}
where the factor $1/2$ associated with the $g_1$ coefficient is
due to the fact that only one hadron beam is involved in the SIDIS
processes, in contrast to two hadron beams in the Drell-Yan type
processes. Furthermore, the additional $g_h$ term is to parametrize
the non-perturbative effect associated with the fragmentation of
the final state parton into the observed hadron.
$z_h$ represents the momentum fraction of the virtual photon
carried by the final state hadron in the SIDIS process.
Using the findings from fitting to the Drell-Yan type data for the
3 shape parameters ($g_1, g_2, g_3$), we found that
the experimental data from HERMES and COMPASS
can be well described by the SIYY-2 form with
\begin{eqnarray}
g_h=0.042 \ .
\end{eqnarray}
Here, we are not performing a fit for the lack of more precise data.
Instead, we merely find a value of $g_h$ to show that the proposed
SIYY-2 form can describe the existing SIDIS data if only the leading
power prediction (defined as the resummation result without including
the $Y$-term) is used for the comparison, cf. Fig. 9.
The reason for not including the $Y$-term in this comparison is
that the typical energy scales ($Q^2$) of the
 SIDIS data from HERMES and COMPASS experiments
are low, at a few GeV.
Hence, the theoretical uncertainties in
applying the CSS formalism is not well under control,
and the $Y$-term contribution is expected to be sizable
as compared to the leading power contribution.
This is illustrated in Fig. 10.
Followed by that, we showed in Fig. 11 that for future
SIDIS data with a larger $Q^2$ value, the CSS formalism will
provide a better description of the data, where the
$Y$-term contribution is expected to be small in the
region that the resummation effect is important, {\it i.e.,}
in the low transverse momentum region.
In other words, we have demonstrated that the proposed
SIYY-1 and SIYY-2 non-perturbative
forms can be used in the CSS resummation formalism
to simultaneously describe the
Drell-Yan and SIDIS data.

Since the $Q^2$ dependence in the non-perturbative functions is
universal among the spin-independent and spin-dependent observables
in the hard scattering processes, including Drell-Yan
lepton pair production in hadronic collisions, semi-inclusive hadron
production in DIS, and di-hadron production in $e^+e^-$ annihilations,
we expect that the new function obtained in this paper shall have
broad applications in the analysis of the spin asymmetries
in these processes. One particular example is the so-called Sivers
single transverse spin asymmetries in SIDIS and Drell-Yan
processes, where the sign change of the asymmetries in these
two processes has been one of top questions in hadron physics.
With the proposed SIYY-1 and SIYY-2 forms, we could further test
the universality property of the TMD formalism.

Before concluding this section, we would like to update the
result of the fit using a pure Gaussian form, similar to the
BLNY or KN fits, but including the more precise $Z$ boson 
data from the CDF and D0 Collaborations at the Tevatron Run II. 
As noted in the Introduction section, it is
difficult to simultaneously describe the
Drell-Yan and SIDIS data using a pure Gaussian form.
Nevertheless, it is still useful to present an update of
the type of fit which
is found to be able to describe very well the Drell-Yan type data
such as the production of $W$ and $Z$ bosons at the Tevatron
and the LHC.
We will name this updated pure Gaussian form
as the SIYY-g form here, which is
\begin{equation}
S^{\rm SIYY-g}_{NP}=
g_1b^2+g_2b^2\ln\left({Q}/{2 Q_0}\right)+g_3b^2\ln(100 x_1x_2)  \ ,
\end{equation}
for describing only the Drell-Yan type of processes in hadronic collisions.
After fixing $Q_0$ to be $1.55$\,GeV and $b_{max}=1.5\,{\rm GeV}^{-1}$,
we found that at the 68\%~C.L.,
\begin{eqnarray}
g_1&=&0.181\pm 0.005,\nonumber \\
g_2&=&0.167\pm 0.01,\nonumber \\
g_3&=& 0.003 \ ,
\end{eqnarray}
where we have fixed $g_3$ at its best fit value.
The quality of the fit to the same set of Drell-Yan data
is similar to that using the SIYY-1 form.
The obtained $g_2$ value is consistent with the
estimation from lattice QCD calculation,
related to the vacuum average of the Wilson loop operator,
as $0.19^{+0.12}_{-0.09}\, {\rm GeV}^2$~\cite{Tafat:2001in}.
As noted before, in the small $b$ region (much less than
$b_{max}$), $\ln (b/b_*) \sim {b^2 /(2 b^2_{max})} $.
Clearly, the value of $g_2$ found in the SIYY-g fit is
consistent with our findings in the SIYY-1 fit whose
$g_2$ value in the small $b$ limit
corresponds to $0.84/(2*1.5^2)=0.187$.

\section{acknowledgements }

We thank Alexei Prokudin for cross checking the $Y$-term
contributions in SIDIS in Sec. IV.
This work was partially supported by the U. S. Department of Energy via grant DE-AC02-05CH11231
and by the U.S. National Science Foundation under Grant No. PHY-1417326.

{\it Note Added}: After this paper was finished, we noticed a preprint of Ref.~\cite{Boglione:2014oea},
which also studied the $Y$-term and matching between the resummation and collinear
calculations. Their conclusion is consistent with ours.

\end{document}